\documentclass[10pt,a4paper,english,twocolumn, journal]{IEEEtran}
\usepackage[T1]{fontenc}
\usepackage[latin9]{inputenc}
\usepackage{array}
\usepackage{verbatim}
\usepackage{float}
\usepackage{calc}
\usepackage{amsmath}
\usepackage{amsthm}
\usepackage{amssymb}
\usepackage{stackrel}
\usepackage{graphicx}
\usepackage{setspace}

\makeatletter

\providecommand{\tabularnewline}{\\}
\floatstyle{ruled}
\newfloat{algorithm}{tbp}{loa}
\providecommand{\algorithmname}{Algorithm}
\floatname{algorithm}{\protect\algorithmname}

\theoremstyle{plain}
\newtheorem{thm}{\protect\theoremname}
\theoremstyle{plain}
\newtheorem{lem}[thm]{\protect\lemmaname}
\theoremstyle{remark}
\newtheorem{rem}[thm]{\protect\remarkname}
\theoremstyle{plain}
\newtheorem{cor}[thm]{\protect\corollaryname}
\theoremstyle{definition}
\newtheorem{defn}[thm]{\protect\definitionname}

\usepackage{subfigure}
\usepackage{epstopdf}
\usepackage{cite}
\usepackage{balance}

\makeatother

\usepackage{babel}
\providecommand{\corollaryname}{Corollary}
\providecommand{\definitionname}{Definition}
\providecommand{\lemmaname}{Lemma}
\providecommand{\remarkname}{Remark}
\providecommand{\theoremname}{Theorem}

\begin{document}

\title{Dense Small Cell Networks: From Noise-Limited to Dense Interference-Limited}

\author{\noindent {\normalsize{}Bin Yang, }\textit{\normalsize{}Student Member,
IEEE}{\normalsize{}, Guoqiang Mao, }\textit{\normalsize{}Fellow}\emph{\normalsize{},
IEEE,}{\normalsize{} Ming Ding, }\textit{\normalsize{}Member, IEEE,}{\normalsize{}
Xiaohu Ge, }\textit{\normalsize{}Senior Member, IEEE, }\textit{\emph{\normalsize{}Xiaofeng
Tao,}}\textit{\normalsize{} Senior Member, IEEE}\thanks{Manuscript received June 29, 2017; revised October 22, 2017 and December
14, 2017; accepted December 19, 2017. Date of publication xx xx, xx;
date of current version xx xx, xx. The authors would like to acknowledge
the support from the NFSC Major International Joint Research Project
under the grant 61210002, National Natural Science Foundation of China
(NSFC) under the grants 61301128 and 61461136004, the Ministry of
Science and Technology (MOST) of China under the grants 2014DFA11640
and 2012DFG12250, the Fundamental Research Funds for the Central Universities
under the grant 2015XJGH011, the Special Research Fund for the Doctoral
Program of Higher Education (SRFDP) under grant 20130142120044. This
research is partially supported by the EU FP7-PEOPLE-IRSES, project
acronym S2EuNet (grant no. 247083), project acronym WiNDOW (grant
no. 318992), project acronym CROWN (grant no. 610524) and the China
Scholarship Council (CSC).}\thanks{Copyright (c) 2015 IEEE. Personal use of this material is permitted.
However, permission to use this material for any other purposes must
be obtained from the IEEE by sending a request to pubs-permissions@ieee.org.}\thanks{Correspondence author: Prof. Xiaohu Ge (Email: xhge@mail.hust.edu.cn,
Tel: +86-27-87557941 ext 822).}\thanks{Bin Yang and Xiaohu Ge are with the School of Electronic Information
and Communications, Huazhong University of Science and Technology,
China (Email: yangbin@hust.edu.cn, xhge@mail.hust.edu.cn). }\thanks{Guoqiang Mao is with the School of Computing and Communication, The
University of Technology Sydney, Australia (e-mail: g.mao@ieee.org). }\thanks{Ming Ding is with Data61, CSIRO, Australia (Email: ming.ding@data61.csiro.au). }\thanks{Xiaofeng Tao is with the National Engineering Lab. for Mobile Network
Security, Beijing University of Posts and Telecommunications, China.
(Email: taoxf@bupt.edu.cn). }\thanks{Part of this paper was presented at the IEEE International Conference
on Communications, Paris, France, May 2017 \cite{Yang17PerformanceC}.}}
\maketitle
\begin{abstract}
Considering both non-line-of-sight (NLoS) and line-of-sight (LoS)
transmissions, the transitional behaviors from noise-limited regime
to dense interference-limited regime have been investigated for the
fifth generation (5G) small cell networks (SCNs). Besides, we identify
four performance regimes based on base station (BS) density, i.e.,
\emph{(i) the noise-limited regime, (ii) the signal-dominated regime,
(iii) the interference-dominated regime, and (iv) the interference-limited
regime}. To characterize the performance regime, we propose a unified
framework analyzing the future 5G wireless networks over generalized
shadowing/fading channels, in which the user association schemes based
on the strongest instantaneous received power (SIRP) and the strongest
average received power (SARP) can be studied, while NLoS/LoS transmissions
and multi-slop path loss model are considered. Simulation results
indicate that different factors, i.e., noise, desired signal, and
interference, successively and separately dominate the network performance
with the increase of BS density. Hence, our results shed new light
on the design and management of SCNs in urban and rural areas with
different BS deployment densities. 
\end{abstract}


\section{Introduction}

According to the study of Prof. Webb~\cite{Webb2007Wireless,LopezPerez15Towards},
the wireless capacity has increased about 1 million fold from 1950
to 2000. Data shows that around $2700\times$ improvement was achieved
by cell splitting and network densification, while the rest of the
gain, was mainly obtained from the use of a wider spectrum, better
coding techniques, and modulation schemes. In this context, \emph{network
densification} has been and will still be the main force to achieve
the $1000\times$ fold increase of data rates in the future fifth
generation (5G) wireless networks~\cite{Andrews14What,Cisco16Visual},
due to its large spectrum reuse as well as its easy management. In
this paper, we focus on the analysis of transitional behaviors for
small cell networks (SCNs) using an orthogonal deployment with the
existing macrocells, i.e., small cells and macrocells are operating
on different frequency spectrum~\cite{Ge165G,Ge15Energy,Yang16Coverage,Yang15A}.

Regarding the network performance of SCNs, a fundamental question
is: \emph{What is the performance trend of SCNs as the base station
(BS) density increases?} In this paper, we answer this question and
identify four performance regimes based on BS density with considerations
of non-line-of-sight (NLoS) and line-of-sight (LoS) transmissions.
These four performance regimes are: (i) the noise-limited regime,
(ii) the signal NLoS-to-LoS-transition regime, (iii) the interference
NLoS-to-LoS-transition regime, and (iv) the dense interference-limited
regime. To characterize the performance regime, we propose a unified
framework analyzing the future 5G wireless networks over generalized
shadowing/fading channels. %
The main contributions of this paper are summarized as follows:
\begin{itemize}
\item We reveal the transitional behaviors from noise-limited regime to
dense interference-limited regime in SCNs and analyze in detail the
factors that affect the performance trend. The analysis results will
benefit the design and management of SCNs in urban and rural areas
with different BS deployment densities.
\item We identify four performance regimes based on BS density. For the
discovered regimes, we present tractable definitions for the regime
boundaries. More specifically, 
\begin{itemize}
\item The boundary between the noise-limited regime and the signal NLoS-to-LoS-transition
regime;
\item The boundary between the signal-dominated regime and the interference
NLoS-to-LoS-transition regime;
\item The boundary between the interference-dominated regime and the interference-limited
regime.
\end{itemize}
\item An accurate SCN model and generalized theoretical analysis: For characterizing
the NLoS-to-LoS transitional behaviors in SCNs, we propose a unified
framework, in which the user association strategies based on the strongest
instantaneous received power (SIRP) and the strongest average received
power (SARP) can be studied, assuming generalized shadowing/fading
channels, multi-slop path loss model and incorporating both NLoS and
LoS transmissions.
\end{itemize}

The remainder of this paper is organized as follows. In Section \ref{sec:Motivation-and-Related},
motivations and some recent work closely related to ours are presented.
Section \ref{sec:System-Model} introduces the system model and network
assumptions. An important theorem used in the analysis on transforming
the original network into an equivalent distance-dependent network,
i.e., the Equivalence Theorem, is presented and proven in Section
\ref{sec:Equivalence-Theorem-and}. Section \ref{sec:SINR-Coverage-Probability}
studies the coverage probability and the ASE of SCNs, more specifically,
several special cases are also investigated. In Section \ref{sec:Simulations},
the analytical results are validated via Monte Carlo simulations.
Besides, the transitional behaviors are elaborated and tractable definitions
for the regime boundaries are presented. Finally, Section \ref{sec:Conclusions-and-Future}
concludes this paper and discusses possible future work.

\section{\label{sec:Motivation-and-Related}Motivations and Related Work}

The modeling of the spatial distribution of SCNs using stochastic
geometry has resulted in significant progress in understanding the
performance of cellular networks \cite{Andrews11A,Blaszczyszyn13Using,Dhillon14Downlink}.
Random spatial point processes, especially the homogeneous Poisson
point process (PPP), have now been widely used to model the locations
of small cell BSs in various scenarios. Existing results are likely
to analyze the performance assuming that the networks operate in the
noise-limited regime or the interference-limited regime. However,
the transitional behaviors from noise-limited regime to interference-limited
regime were rarely mentioned in their work. Some assumptions in the
system model were even conflicted with each other, e.g., in \cite{Singh15Tractable}
and \cite{Bai15Coverage}, the millimeter wave networks were assumed
to be noise-limited and interference-limited, respectively. Besides,
most work is usually based on certain simplified assumptions, e.g.,
Rayleigh fading, a single path loss exponent with no thermal noise,
etc, for analytical tractability, which may not hold in a more realistic
scenario. For instance, consider a SCN in urban areas, the path loss
model may not follow a single power law relationship in the near-filed
and thus non-singular \cite{Liu17Effect,Khamesi16Energy} or multiple-slop
path loss model \cite{Zhang15Downlink} should be applied. Besides,
signal transmissions between BSs and MUs are frequently affected by
reflection, diffraction, and even blockage due to high-rise buildings
in urban areas, and thus NLoS/LoS transmissions should also be considered
\cite{Bai15Coverage}. As a consequence, the detailed analysis of
transitional behaviors are needed, with considerations of a more generalized
propagation model incorporating both NLoS and LoS transmissions, to
cope with these new characteristics in SCNs.

A number of more recent work had a new look at dense SCNs considering
more practical propagation models. The closest system model to the
one in this paper are in \cite{ShokriGhadikolaei16The,Ding16Performance,Bai15Coverage,DiRenzo15StochasticJ,Singh15Tractable,Arnau16Impact,Liu17Effect,DiRenzo16The,Renzo13Average}.
In \cite{ShokriGhadikolaei16The}, the transitional behaviors of interference
in millimeter wave networks was analyzed, but it focused on the medium
access control. In \cite{Bai15Coverage} and \cite{Ding16Performance},
the coverage probability and capacity were calculated based on the
smallest path loss cell association model assuming multi-path fading
modeled as Rayleigh fading and Nakagami-$m$ fading, respectively.
However, shadowing was ignored in their models, which may not be very
practical for a SCN. The authors of \cite{Singh15Tractable} and \cite{DiRenzo15StochasticJ}
analyzed the coverage and capacity performance in millimeter wave
cellular networks. In \cite{Singh15Tractable}, self-backhauled millimeter
wave cellular networks were analyzed assuming a cell association scheme
based on the smallest path loss. In \cite{DiRenzo15StochasticJ},
a three-state statistical model for each link was assumed, in which
a link can either be in a NLoS, LoS or an outage state. Besides, both
\cite{Singh15Tractable} and \cite{DiRenzo15StochasticJ} assumed
a noise-limited network ignoring inter-cell interference, which may
not be very practical since modern wireless networks generally work
in an interference-limited region. In \cite{Arnau16Impact}, the authors
assumed Rayleigh fading for NLoS transmissions and Nakagami-$m$ fading
for LoS transmissions which is more practical than work in \cite{Ding16Performance}.
However, the cell association scheme in \cite{Arnau16Impact} is only
applicable to the scenario where the SINR threshold is greater than
0 dB. Besides, the ASE performance was not analyzed in \cite{Arnau16Impact}.
In \cite{Liu17Effect}, a near-filed path loss model with bounded
path loss was studied. In \cite{DiRenzo16The}, a tractable performance
evaluation method, i.e., the intensity matching, was proposed to model
and optimize the networks. Renzo \emph{et al.} \cite{Renzo13Average}
also introduced an analytical framework based on the strongest average
received signal power associations scheme which is applicable to general
fading distributions, including composite fading channels, to analyze
the average rate of heterogeneous networks using a single-slope path
loss model.

To summarize, in this paper, we propose a more generalized framework
to analyze the \emph{transitional behaviors} for SCNs compared with
the work in \cite{ShokriGhadikolaei16The,Ding16Performance,Bai15Coverage,DiRenzo15StochasticJ,Singh15Tractable,Arnau16Impact,Liu17Effect,DiRenzo16The}.
Our framework takes into account a cell association scheme based on
the strongest received signal power, probabilistic NLoS and LoS transmissions,
multi-slop path loss model, multi-path fading and/or shadowing. Furthermore,
the proposed framework can also be applied to analyze dense SCNs,
where BSs are distributed according to non-homogeneous PPPs, i.e.,
the BS density is spatially varying.

\section{\label{sec:System-Model}System Model}

We consider a homogeneous SCN in urban areas and focus on the analysis
of downlink performance. We assume that BSs are spatially distributed
on an infinite plane and the locations of BSs $\boldsymbol{X}_{i}$
follow a homogeneous PPP denoted by $\Phi=\left\{ \boldsymbol{X}_{i}\right\} $
with an density of $\lambda$, where $i$ is the BS index \cite{Ge14Performance}.
MUs are deployed according to another independent homogeneous PPP
denoted by $\Phi_{\textrm{u}}$ with an density of $\lambda_{\textrm{u}}$.
All BSs in the network operate at the same power $P_{\textrm{t}}$
and share the same bandwidth. Within a cell, MUs use orthogonal frequencies
for downlink transmissions and therefore \emph{intra-cell interference}
is not considered in our analysis. However, adjacent BSs may generate
\emph{inter-cell interference} to MUs, which is the primary focus
of our work.

\subsection{\label{subsec:Signal-Propagation-Model}Path Loss Model}

In a downlink SCN, the long-distance signal attenuation is modeled
by a monotone, non-increasing and continuous path loss function $l\left(R_{i}\right):\left[0,\infty\right]\mapsto\left[0,\infty\right]$
and $l\left(R_{i}\right)$ decays to zero asymptotically, where $R_{i}=\left\Vert \boldsymbol{X}_{i}\right\Vert $
denotes the Euclidean distance between a BS at $\boldsymbol{X}_{i}$
and the typical MU (aka the probe MU or the tagged MU) located at
the origin $o$. Specifically, a multi-slop path loss function \cite{Zhang15Downlink,Ding16Performance}
is utilized in which the distance $R_{i}$ is segmented into $N$
pieces. Compared with the single-slope path loss model, the multi-slope
path loss model is more flexible and can characterize the future networks
instead of only depending on the existing cellular works. Besides,
the standard path loss model does not accurately capture the dependence
of the path loss exponent $\alpha$ on the link distance in many important
situations \cite{Zhang15Downlink,Ding16Performance}. The multi-slop
path loss function is written as
\begin{equation}
l\left(R_{i}\right)=\begin{cases}
l_{1}\left(R_{i}\right), & \textrm{when }0\leqslant R_{i}\leqslant d_{1}\\
l_{2}\left(R_{i}\right), & \textrm{when }d_{1}<R_{i}\leqslant d_{2}\\
\vdots & \vdots\\
l_{N}\left(R_{i}\right), & \textrm{when }R_{i}>d_{N-1}
\end{cases},\label{eq:prop_PL_model}
\end{equation}
where each piece $l_{n}\left(R_{i}\right),n\in\left\{ 1,2,\ldots,N\right\} \triangleq\mathcal{N}$
incorporates both NLoS and LoS transmissions, whose performance impact
is attracting growing interest among researchers recently. In reality,
the occurrence of NLoS or LoS transmissions depends on various environmental
factors, including geographical structure, distance, and clusters,
etc. Note that the corresponding points in each region form independent
point processes denoted by $\Phi_{n},n\in\mathcal{N}$, i.e.,
\begin{equation}
\begin{cases}
\Phi_{1}\triangleq\left\{ \boldsymbol{X}_{i}\left|\left\Vert \boldsymbol{X}_{i}\right\Vert \in\left[0,d_{1}\right]\right.\right\} , & \textrm{when }n=1\\
\Phi_{n}\triangleq\left\{ \boldsymbol{X}_{i}\left|\left\Vert \boldsymbol{X}_{i}\right\Vert \in\left(d_{n-1},d_{n}\right]\right.\right\} , & \textrm{when }n\notin\left\{ 1,N\right\} \\
\Phi_{N}\triangleq\left\{ \boldsymbol{X}_{i}\left|\left\Vert \boldsymbol{X}_{i}\right\Vert \in\left(d_{N-1},\infty\right]\right.\right\} , & \textrm{when }n=N
\end{cases}.
\end{equation}

In the following, we give a simplified one-parameter model of NLoS
and LoS transmissions. The occurrence of NLoS and LoS transmissions
in each piece $l_{n}\left(R_{i}\right)$ can be modeled using probabilities
$p_{n}^{\textrm{NL}}\left(R_{i}\right)$ and $p_{n}^{\textrm{L}}\left(R_{i}\right)$,
respectively, i.e., 
\begin{equation}
l\left(R_{i}\right)\hspace{-0.1cm}=\hspace{-0.1cm}\begin{cases}
\hspace{-0.2cm}\begin{array}{l}
l_{n}^{\textrm{NL}}\left(R_{i}\right),\\
l_{n}^{\textrm{L}}\left(R_{i}\right),
\end{array} & \hspace{-0.2cm}\hspace{-0.3cm}\begin{array}{l}
\textrm{with probability:}~p_{n}^{\textrm{NL}}\left(R_{i}\right)\\
\textrm{with probability:}~p_{n}^{\textrm{L}}\left(R_{i}\right)
\end{array}\hspace{-0.1cm},\end{cases}\label{eq:PL_BS2UE}
\end{equation}
where $l_{n}^{\textrm{NL}}\left(R_{i}\right)$ and $l_{n}^{\textrm{L}}\left(R_{i}\right)$
are the $n$-th piece path loss functions for the NLoS transmission
and the LoS transmission, respectively, $p_{n}^{\textrm{NL}}\left(R_{i}\right)$
and $p_{n}^{\textrm{NL}}\left(R_{i}\right)$ are the probabilities
that the transmissions are NLoS and LoS, respectively, moreover, $p_{n}^{\textrm{NL}}\left(R_{i}\right)$+$p_{n}^{\textrm{L}}\left(R_{i}\right)=1$.

Regarding the mathematical form of $p_{n}^{\textrm{L}}\left(R_{i}\right)$
(or $p_{n}^{\textrm{NL}}\left(R_{i}\right)$), N. Blaunstein~\cite{Blaunstein98Parametric}
formulated $p_{n}^{\textrm{L}}\left(R_{i}\right)$ as a negative exponential
function, i.e., $p_{n}^{\textrm{L}}\left(R_{i}\right)=e^{-\kappa R_{i}}$,
where $\kappa$ is a parameter determined by the density and the mean
length of the blockages lying in the visual path between the typical
MU and BSs. Bai~\cite{Bai14Analysis} extended N. Blaunstein's work
by using random shape theory which shows that $\kappa$ is not only
determined by the mean length but also the mean width of the blockages.
The authors of~\cite{DiRenzo15StochasticJ} and~\cite{Bai14Analysis}
approximated $p_{n}^{\textrm{L}}\left(R_{i}\right)$ by piece-wise
functions and step functions, respectively. Ming \emph{et al.}~\cite{Ding16Performance}
considered $p_{n}^{\textrm{L}}\left(R_{i}\right)$ as a linear function
and a two-piece exponential function, respectively, both recommended
by the 3GPP \cite{3GPP36828,3GPPSpatial2003}. 

It should be noted that the occurrence of NLoS (or LoS) transmissions
is assumed to be independent for different BS-MU pairs. Though such
assumption might not be entirely realistic, e.g., NLoS transmissions
for nearby MUs caused by a large obstacle may be spatially correlated,
the authors of~\cite{Bai14Analysis} showed that the impact of the
independence assumption on the SINR analysis is negligible.

In general, NLoS and LoS transmissions incur different path losses,
which are formulated by\footnote{As the derivations in scenarios with log-normal shadowing is much
more complicated than that with Rayleigh fading, we choose to take
the former as an example. It is found in Eq. (\ref{eq:Power_N}) and
Eq. (\ref{eq:Power_L}) that the model can also be applied to Rayleigh
fading and other generalized shadowing/fading models.}
\begin{equation}
PL_{\textrm{dB},n}^{\textrm{NL}}=A_{\textrm{dB},n}^{\textrm{NL}}+\alpha_{n}^{\textrm{NL}}10\log_{10}R_{i}+\xi_{\textrm{dB},n}^{\textrm{NL}},
\end{equation}
and
\begin{equation}
PL_{\textrm{dB},n}^{\textrm{L}}=A_{\textrm{dB},n}^{\textrm{L}}+\alpha_{n}^{\textrm{L}}10\log_{10}R_{i}+\xi_{\textrm{dB},n}^{\textrm{L}},
\end{equation}
where the path loss is expressed in dB unit, $A_{\textrm{dB},n}^{\textrm{NL}}$
and $A_{\textrm{dB},n}^{\textrm{L}}$ are the $n$-th piece path losses
at the reference distance (usually at 1 meter), $\alpha_{n}^{\textrm{NL}}$
and $\alpha_{n}^{\textrm{L}}$ are respectively the $n$-th piece
path loss exponents for NLoS and LoS transmissions, $\xi_{\textrm{dB},n}^{\textrm{NL}}$
and $\xi_{\textrm{dB},n}^{\textrm{L}}$ are independent Gaussian random
variables with zero means, i.e., $\xi_{\textrm{dB},n}^{\textrm{NL}}\sim\mathcal{N}\left(0,\left(\sigma_{n}^{\textrm{NL}}\right)^{2}\right)$
and $\xi_{\textrm{dB},n}^{\textrm{L}}\sim\mathcal{N}\left(0,\left(\sigma_{n}^{\textrm{L}}\right)^{2}\right)$,
reflecting the signal attenuation caused by shadow fading. The corresponding
model parameters can be found in~\cite{3GPP36828,Mao06Online,Mao09Graph,Mao13Road}.

Accordingly, the $n$-th piece received signal power for NLoS and
LoS transmissions in W (watt) can be respectively expressed by
\begin{equation}
P_{i,n}^{\textrm{NL}}=P_{\textrm{t}}\cdot10^{-A_{\textrm{dB},n}^{\textrm{NL}}/10}\mathcal{H}_{i,n}^{\textrm{NL}}\left(R_{i}\right)^{-\alpha_{n}^{\textrm{NL}}}=B_{n}^{\textrm{NL}}\mathcal{H}_{i,n}^{\textrm{NL}}l_{n}^{\textrm{NL}}\left(R_{i}\right),\label{eq:Power_N}
\end{equation}
and
\begin{equation}
P_{i,n}^{\textrm{L}}=P_{\textrm{t}}\cdot10^{-A_{\textrm{dB},n}^{\textrm{L}}/10}\mathcal{H}_{i,n}^{\textrm{L}}\left(R_{i}\right)^{-\alpha_{n}^{\textrm{L}}}=B_{n}^{\textrm{L}}\mathcal{H}_{i,n}^{\textrm{L}}l_{n}^{\textrm{L}}\left(R_{i}\right),\label{eq:Power_L}
\end{equation}
where $\mathcal{H}_{i,n}^{\textrm{NL}}=\exp\left(\beta\xi_{\textrm{dB},n}^{\textrm{NL}}\right)$
(or $\mathcal{H}_{i,n}^{\textrm{L}}=\exp\left(\beta\xi_{\textrm{dB},n}^{\textrm{L}}\right)$
) denotes log-normal shadowing for NLoS (or LoS) transmission, and
$B_{n}^{\textrm{NL}}=P_{\textrm{t}}\cdot10^{-A_{\textrm{dB},n}^{\textrm{NL}}/10}$,
$B_{n}^{\textrm{L}}=P_{\textrm{t}}\cdot10^{-A_{\textrm{dB},n}^{\textrm{L}}/10}$
and $\beta=-\ln10/10$ are all constants. Note that usually it is
assumed that shadowing among different BS-MU pairs are mutually independent
and identically distributed (i.i.d.) and also independent of BS locations
\cite{Dhillon14Downlink,Andrews11A}, thus $\mathcal{H}_{i,n}^{\textrm{NL}}$
and $\mathcal{H}_{i,n}^{\textrm{L}}$ can be denoted as $\mathcal{H}_{n}^{\textrm{NL}}$
and $\mathcal{H}_{n}^{\textrm{L}}$, respectively, for the the convenience
of expression. Moreover, if we replace $\mathcal{H}_{n}^{\textrm{NL}}$
(or $\mathcal{H}_{n}^{\textrm{L}}$ ) by multi-path fading, i.e.,
$h_{n}^{\textrm{NL}}$ (or $h_{n}^{\textrm{L}}$ ) the model can also
be applied.

Therefore, the received power by the typical MU from BS $\boldsymbol{X}_{i}$
is given by Eq. (\ref{eq:P_received}):
\begin{algorithm*}
\begin{singlespace}
\noindent 
\begin{equation}
P_{i}\left(R_{i}\right)=\begin{cases}
P_{i,1}\left(R_{i}\right)=\begin{cases}
\begin{array}{l}
P_{i,1}^{\textrm{NL}}\left(R_{i}\right)=B_{1}^{\textrm{NL}}\mathcal{H}_{i,1}^{\textrm{NL}}l_{1}^{\textrm{NL}}\left(R_{i}\right),\\
P_{i,1}^{\textrm{L}}\left(R_{i}\right)=B_{1}^{\textrm{L}}\mathcal{H}_{i,1}^{\textrm{L}}l_{1}^{\textrm{L}}\left(R_{i}\right),
\end{array} & \hspace{-0.3cm}\begin{array}{l}
\textrm{with probability: }p_{1}^{\textrm{NL}}\left(R_{i}\right)\\
\textrm{with probability: }p_{1}^{\textrm{L}}\left(R_{i}\right)
\end{array}\end{cases}\hspace{-0.3cm}, & \hspace{-0.3cm}\textrm{when }0\leqslant R_{i}\leqslant d_{1}\\
P_{i,2}\left(R_{i}\right)=\begin{cases}
\begin{array}{l}
P_{i,2}^{\textrm{NL}}\left(R_{i}\right)=B_{2}^{\textrm{NL}}\mathcal{H}_{i,2}^{\textrm{NL}}l_{2}^{\textrm{NL}}\left(R_{i}\right),\\
P_{i,2}^{\textrm{L}}\left(R_{i}\right)=B_{2}^{\textrm{L}}\mathcal{H}_{i,2}^{\textrm{L}}l_{2}^{\textrm{L}}\left(R_{i}\right),
\end{array} & \hspace{-0.3cm}\begin{array}{l}
\textrm{with probability: }p_{2}^{\textrm{NL}}\left(R_{i}\right)\\
\textrm{with probability: }p_{2}^{\textrm{L}}\left(R_{i}\right)
\end{array}\end{cases}\hspace{-0.3cm}, & \hspace{-0.3cm}\textrm{when }d_{1}<R_{i}\leqslant d_{2}\\
\vdots & \vdots\\
P_{i,N}\left(R_{i}\right)=\begin{cases}
\begin{array}{l}
P_{i,N}^{\textrm{NL}}\left(R_{i}\right)=B_{N}^{\textrm{NL}}\mathcal{H}_{i,N}^{\textrm{NL}}l_{N}^{\textrm{NL}}\left(R_{i}\right),\\
P_{i,N}^{\textrm{L}}\left(R_{i}\right)=B_{N}^{\textrm{L}}\mathcal{H}_{i,N}^{\textrm{L}}l_{N}^{\textrm{L}}\left(R_{i}\right),
\end{array} & \hspace{-0.3cm}\begin{array}{l}
\textrm{with probability: }p_{N}^{\textrm{NL}}\left(R_{i}\right)\\
\textrm{with probability: }p_{N}^{\textrm{L}}\left(R_{i}\right)
\end{array}\end{cases}\hspace{-0.3cm}, & \hspace{-0.3cm}\textrm{when }R_{i}>d_{N-1}
\end{cases}.\label{eq:P_received}
\end{equation}
\end{singlespace}
\end{algorithm*}

Based on the path loss model discussed above, for downlink transmissions,
the SINR experienced by the typical MU associated with BS $\boldsymbol{X}_{i}$
can be written as
\begin{align}
\textrm{SINR}_{i} & =\frac{S}{I+\eta}=\frac{P_{i}\left(R_{i}\right)}{\underset{\boldsymbol{X}_{z}\in\Phi\setminus\boldsymbol{X}_{i}}{\sum}P_{z}\left(R_{z}\right)+\eta},\label{eq:SINR}
\end{align}
where $\Phi\setminus\boldsymbol{X}_{i}$ is the Palm point process~\cite{Chiu13Stochastic}
representing the set of interfering BSs in the network to the typical
MU and $\eta$ denotes the noise power at the MU side, which is assumed
to be the additive white Gaussian noise (AWGN). For clarity, we summarize
the notation used in Table \ref{tab:Notation-Summary} for quick access.

\begin{table}
\caption{\label{tab:Notation-Summary}Notation and Simulation Parameters Summary}
\begin{tabular}{|l|>{\raggedright}m{4.5cm}|>{\raggedright}m{1.6cm}|}
\hline 
Notation & Explanation & Value (if applicable)\tabularnewline
\hline 
\hline 
$\Phi$, $\lambda$ & Homogeneous BS PPP and its density & \tabularnewline
\hline 
$\Phi_{n}^{\textrm{NL}}$, $\Phi_{n}^{\textrm{L}}$ & NLoS BS PPP and LoS BS PPP, $\Phi_{n}=\Phi_{n}^{\textrm{NL}}\cup\Phi_{n}^{\textrm{L}}$ & \tabularnewline
\hline 
$\overline{\Phi_{n}^{\textrm{NL}}}$, $\overline{\Phi_{n}^{\textrm{L}}}$ & Equivalent NLoS BS PPP and equivalent LoS BS PPP & \tabularnewline
\hline 
$P_{t}$ & BS transmission power & 30 dBm\tabularnewline
\hline 
$\mathcal{H}_{n}^{\textrm{NL}}$, $\mathcal{H}_{n}^{\textrm{L}}$  & Log-normal shadowing for NLOS and LOS transmissions & \tabularnewline
\hline 
$A_{n}^{\textrm{NL}}$ , $A_{n}^{\textrm{L}}$  & Path loss at the the reference distance (1m) & 30.8, 2.7 \cite{3GPP36828}\tabularnewline
\hline 
$\sigma_{n}^{\textrm{NL}}$, $\sigma_{n}^{\textrm{L}}$ & Standard deviation of shadowing for NLoS and LoS transmissions & 4 dB, 3 dB \cite{3GPP36828}\tabularnewline
\hline 
$\mu_{n}^{\textrm{NL}}$, $\mu_{n}^{\textrm{L}}$ & Rate of Rayleigh fading for NLoS and LoS transmissions & 1, 1\tabularnewline
\hline 
$\alpha_{n}^{\textrm{NL}}$, $\alpha_{n}^{\textrm{L}}$ & Path loss exponents for NLoS and LoS transmissions & 4.28, 2.42 \cite{3GPP36828}\tabularnewline
\hline 
$\eta$ & Noise power & -95 dBm \cite{3GPP36828}\tabularnewline
\hline 
$\overline{R_{i,n}^{\textrm{NL}}}$, $\overline{R_{i,n}^{\textrm{L}}}$ & Equivalent distance for NLoS and LoS transmissions & \tabularnewline
\hline 
$\Lambda_{n}^{\textrm{NL}}$, $\Lambda_{n}^{\textrm{L}}$  & Intensity measure of $\overline{\Phi^{\textrm{NL}}}$ and $\overline{\Phi^{\textrm{L}}}$ & \tabularnewline
\hline 
$\lambda_{n}^{\textrm{NL}}$, $\lambda_{n}^{\textrm{L}}$ & Intensity of $\overline{\Phi^{\textrm{NL}}}$ and $\overline{\Phi^{\textrm{L}}}$ & \tabularnewline
\hline 
$d$ & Radius of LoS region & 250 m \cite{Bai15Coverage,Singh15Tractable}\tabularnewline
\hline 
$T$ & SINR (or SIR) threshold & 0 dB\tabularnewline
\hline 
$I$, $I^{\textrm{NL}}$, $I^{\textrm{L}}$ & Aggregate interference, aggregate interference from NLoS and LoS transmissions & \tabularnewline
\hline 
\end{tabular}
\end{table}

\subsection{Cell Association Scheme}

Considering NLoS and LoS transmissions, two cell association schemes
can be studied, based on the strongest average received power and
the strongest instantaneous SINR, respectively. As for the strongest
instantaneous SINR association, the typical MU associates itself to
the BS $\boldsymbol{X}_{i}^{*}$ given by
\begin{align}
\boldsymbol{X}_{i}^{*} & =\arg\underset{\boldsymbol{X}_{i}\in\Phi}{\max}\left\{ \textrm{SINR}_{i}\right\} .\label{eq:maxSINR}
\end{align}
Intuitively, the strongest instantaneous SINR association is equivalent
to the strongest instantaneous received signal power association.
Such intuition is formally presented and proved in Lemma~\ref{lem: lem1}.
\begin{lem}
\label{lem: lem1}For a non-negative set $\Xi=\left\{ a_{q}\right\} $,
$q\in\mathbb{N}$, $\frac{a_{m}}{\underset{q\neq m}{\sum}a_{q}+W}>\frac{a_{n}}{\underset{q\neq n}{\sum}a_{q}+W}$
if and only if $a_{m}>a_{n}$, $\forall a_{m},a_{n}\in\Xi$.
\end{lem}
\begin{IEEEproof}
For a non-negative set $\Xi=\left\{ a_{q}\right\} $, $q\in\mathbb{N}$,
$\frac{a_{m}}{\underset{q}{\sum}a_{q}+W}>\frac{a_{n}}{\underset{q}{\sum}a_{q}+W}$
if and only if $a_{m}>a_{n}$, thus $\frac{a_{m}}{\underset{q}{\sum}a_{q}+W-a_{m}}>\frac{a_{n}}{\underset{q}{\sum}a_{q}+W-a_{n}}$
if and only if $a_{m}>a_{n}$, which completes the proof.
\end{IEEEproof}
Lemma \ref{lem: lem1} states that providing the strongest instantaneous
SINR is equivalent to providing the strongest instantaneous received
power to the typical MU. It follows from Eq. (\ref{eq:maxSINR}) and
Lemma \ref{lem: lem1} that the BS associated with the typical MU
can also be written as
\begin{align}
\left(\boldsymbol{X}_{i},\textrm{U},\mathcal{N}\right)^{*} & =\arg\underset{(\boldsymbol{X}_{i},\textrm{U},\mathcal{N})\in\mathbb{S}}{\max}\left\{ B_{n}^{\textrm{U}}h_{n}^{\textrm{U}}\left(R_{i}\right)^{-\alpha_{n}^{\textrm{U}}}\right\} ,\label{eq:inspower}
\end{align}
where $\boldsymbol{X}_{i}\in\Phi$, $\textrm{U}\in\left\{ \textrm{NL},\textrm{L}\right\} $
and the set $\mathbb{S}=\Phi\times\left\{ \textrm{NL},\textrm{L}\right\} \times\mathcal{N}$.
Note that under SIRP, we ignore shadowing, i.e., $\mathcal{H}_{n}^{\textrm{U}}$,
for the sake of simplicity.

As for the SARP, the typical MU associates itself to the BS $\left(\boldsymbol{X}_{i},\textrm{U},\mathcal{N}\right)^{*}$
given by
\begin{align}
\left(\boldsymbol{X}_{i},\textrm{U},\mathcal{N}\right)^{*} & =\arg\underset{(\boldsymbol{X}_{i},\textrm{U},\mathcal{N})\in\mathbb{S}}{\max}\left\{ B_{n}^{\textrm{U}}\mathcal{H}_{n}^{\textrm{NL}}\left(R_{i}\right)^{-\alpha_{n}^{\textrm{U}}}\right\} .\label{eq:avepower}
\end{align}
Note that under SARP, we ignore multi-path fading, i.e., $h_{n}^{\textrm{U}}$,
for the sake of simplicity. In the following, both cell association
schemes will be studied to characterize the network performance. 

\section{\label{sec:Equivalence-Theorem-and}The Equivalence of SCNs }

Before presenting our main analytical results, firstly we introduce
the Equivalence Theorem which will be used throughout the paper. The
purpose of introducing the Equivalence Theorem is to unify the analysis
considering different multi-path fading and/or shadowing, and to reduce
the complexity of our theoretical analysis. Then based on this theorem,
we derive the cumulative distribution function (CDF) of the strongest
received signal power.

\subsection{The Equivalence of SCNs}

In this subsection, an equivalent SCN to the one being analyzed will
be introduced, which specifies how the intensity measure and the intensity
are changed after a transformation of original PPPs. Under SARP, denoting
by 
\begin{equation}
\overline{R_{i,n}^{\textrm{NL}}}=R_{i}\cdot\left(B_{n}^{\textrm{NL}}\mathcal{H}_{n}^{\textrm{NL}}\right)^{-1/\alpha_{n}^{\textrm{NL}}}\label{eq:R_N}
\end{equation}
and
\begin{equation}
\overline{R_{i,n}^{\textrm{L}}}=R_{i}\cdot\left(B_{n}^{\textrm{L}}\mathcal{H}_{n}^{\textrm{L}}\right)^{-1/\alpha_{n}^{\textrm{L}}},\label{eq:R_L}
\end{equation}
the received signal power in Eq. (\ref{eq:Power_N}) and Eq. (\ref{eq:Power_L})
can be written as
\begin{align}
P_{i,n}^{\textrm{NL}} & =\left(\overline{R_{i,n}^{\textrm{NL}}}\right)^{-\alpha_{n}^{\textrm{NL}}}\label{eq:P_rec_N}
\end{align}
and
\begin{align}
P_{i,n}^{\textrm{L}} & =\left(\overline{R_{i,n}^{\textrm{L}}}\right)^{-\alpha_{n}^{\textrm{L}}}.\label{eq:P_rec_L}
\end{align}

Note that from the viewpoint of the typical MU, each BS in the infinite
plane $\mathbb{R}^{2}$ is either a NLoS BS or a LoS BS. Accordingly,
we perform a thinning procedure on points in the PPP $\Phi_{n}$ to
model the distributions of NLoS BSs and LoS BSs, respectively. That
is, each BS in $\Phi_{n}$ will be kept if a BS has a NLoS transmission
with the typical MU, thus forming a new point process denoted by $\Phi_{n}^{\textrm{NL}}$
. While BSs in $\Phi_{n}\setminus\Phi_{n}^{\textrm{NL}}$ form another
point process denoted by $\Phi_{n}^{\textrm{L}}$, representing the
set of BSs with LoS path to the typical MU. As a consequence of the
independence assumption between LoS and NLoS transmissions mentioned
above, $\Phi_{n}^{\textrm{NL}}$ and $\Phi_{n}^{\textrm{L}}$ are
two independent non-homogeneous PPPs with intensity\footnote{In this article, density and intensity have the same meaning.}
$\lambda p_{n}^{\textrm{NL}}\left(R_{i}\right)$ and $\lambda p_{n}^{\textrm{L}}\left(R_{i}\right)$,
respectively. 

Through the above transformation which scales the distances between
the typical MU and all other BSs using Eq. (\ref{eq:R_N}) and (\ref{eq:R_L}),
the scaled point process for NLoS BSs (or LoS BSs) still remains a
PPP denoted by $\overline{\Phi_{n}^{\textrm{NL}}}$ (or $\overline{\Phi_{n}^{\textrm{L}}}$
) according to the displacement theorem \cite[Theorem 1.3.9]{Baccelli09Stochastic}.
In other words, $\overline{\Phi_{n}^{\textrm{NL}}}$ (or $\overline{\Phi_{n}^{\textrm{L}}}$
) is obtained by randomly and independently displacing each point
of $\Phi_{n}^{\textrm{NL}}$(or $\Phi_{n}^{\textrm{L}}$ ) to some
new location according to the kernel $p=\Pr\left[\overline{R_{i,n}^{\textrm{NL}}}\in b\left(0,t\right)\right]$
(or $p=\Pr\left[\overline{R_{i,n}^{\textrm{L}}}\in b\left(0,t\right)\right]$
). As the transformation is mutually independent, the new point process
is still a PPP. The detailed proof can be obtained in \cite[Lemma 1]{Blaszczyszyn13Using}
and we omitted it for space limitation. The intuition is that in the
equivalent networks, the received signal power and cell association
scheme are only dependent on the new equivalent distance $\overline{R_{i,n}^{\textrm{NL}}}$
(or $\overline{R_{i,n}^{\textrm{L}}}$ ) between the BSs and the typical
MU, while the effects of transmit power, multi-path fading and shadowing
are incorporated into the equivalent intensity (or the equivalent
intensity measure) of the transformed point process. Besides, $\overline{\Phi_{n}^{\textrm{NL}}}$
and $\overline{\Phi_{n}^{\textrm{L}}}$ are mutually independent because
of the independence between $\Phi_{n}^{\textrm{NL}}$ and $\Phi_{n}^{\textrm{L}}$.
As a result, the performance analysis involving path loss, multi-path
fading, shadowing, etc, can be handled in a unified framework, which
motivates the following theorem.
\begin{thm}[The Equivalence Theorem]
\label{thm: Equivalence theorem} Assume that a general fading or
shadowing satisfy $\mathbb{E}_{\mathcal{H}_{n}^{\textrm{U}}}\left[\left(\mathcal{H}_{n}^{\textrm{U}}\right)^{2/\alpha_{n}^{\textrm{U}}}\right]<\infty$.
The system which consists of two non-homogeneous PPPs with intensities
$\lambda p_{n}^{\textrm{NL}}\left(R_{i}\right)$ and $\lambda p_{n}^{\textrm{L}}\left(R_{i}\right)$
respectively, representing the sets of NLoS and LoS BSs, and in which
each MU is associated with the BS providing the strongest received
signal power is equivalent, in terms of performance to the typical
MU located at the origin, to another system consisting of two non-homogeneous
PPPs with intensities (functions) $\lambda_{n}^{\textrm{NL}}\left(\cdot\right)$
and $\lambda_{n}^{\textrm{L}}\left(\cdot\right)$ respectively, representing
the sets of NLoS and LoS BSs, and in which the typical MU is associated
with the nearest BS. Moreover, intensities (functions) $\lambda_{n}^{\textrm{NL}}\left(\cdot\right)$
and $\lambda_{n}^{\textrm{L}}\left(\cdot\right)$ are respectively
given by
\begin{equation}
\lambda_{n}^{\textrm{NL}}\left(t\right)=\frac{\textrm{d}}{\textrm{d}t}\Lambda_{n}^{\textrm{NL}}\left(\left[0,t\right]\right)\label{eq:Lambda_N}
\end{equation}
and
\begin{equation}
\lambda_{n}^{\textrm{L}}\left(t\right)=\frac{\textrm{d}}{\textrm{d}t}\Lambda_{n}^{\textrm{L}}\left(\left[0,t\right]\right),\label{eq:Lambda_L}
\end{equation}
where 
\begin{equation}
\Lambda_{n}^{\textrm{NL}}\left(\left[0,t\right]\right)=\mathbb{E}_{\mathcal{H}_{n}^{\textrm{NL}}}\left[2\pi\lambda\int_{R_{i}=d_{n-1}}^{R_{i,\max}^{\textrm{NL}}}p_{n}^{\textrm{NL}}\left(R_{i}\right)R_{i}\textrm{d}R_{i}\right]\label{eq:Measure_N}
\end{equation}
and
\begin{equation}
\Lambda_{n}^{\textrm{L}}\left(\left[0,t\right]\right)=\mathbb{E}_{\mathcal{H}_{n}^{\textrm{L}}}\left[2\pi\lambda\int_{R_{i}=d_{n-1}}^{R_{i,\max}^{\textrm{L}}}p_{n}^{\textrm{L}}\left(R_{i}\right)R_{i}\textrm{d}R_{i}\right],\label{eq:Measure_L}
\end{equation}
where $R_{i,\max}^{\textrm{NL}}=\min\left\{ d_{n},t\left(B_{n}^{\textrm{NL}}\mathcal{H}_{n}^{\textrm{NL}}\right)^{1/\alpha_{n}^{\textrm{NL}}}\right\} $
and $R_{i,\max}^{\textrm{L}}=\min\left\{ d_{n},t\left(B_{n}^{\textrm{L}}\mathcal{H}_{n}^{\textrm{L}}\right)^{1/\alpha_{n}^{\textrm{L}}}\right\} $.
\end{thm}
\begin{IEEEproof}
See Appendix A.
\end{IEEEproof}
In \cite{Liu16Optimal}, a similar theorem which was also extended
from Blaszczyszyn's work \cite{Blaszczyszyn13Using,Blaszczyszyn15Studying}
was proposed to analyze a $n$-dimensional network, in which NLoS
and LoS transmissions are not considered. By utilizing the Equivalence
theorem above, the transformed cellular network has the exactly same
performance for the typical MU with respect to the coverage probability
and the ASE compared with the original network, which is proved in
Appendix A and validated by Monte Carlo simulations in Section \ref{sec:Simulations}.
After transformation, the received signal power and cell association
scheme are only dependent on the equivalent distance between the BSs
and the typical MU, i.e., $\overline{R_{i,n}^{\textrm{NL}}}$ and
$\overline{R_{i,n}^{\textrm{L}}}$ , while the effects of transmit
power, multi-path fading (under SIRP) and shadowing (under SARP) are
incorporated into the equivalent intensity shown in Eq. (\ref{eq:Lambda_N})
and Eq. (\ref{eq:Lambda_L}). Therefore, the complexity of theoretical
analysis can be significantly reduced.
\begin{rem}
From Lemma \ref{lem: lem1} and Theorem \ref{thm: Equivalence theorem},
any cell association scheme without considering the status of BSs
and MUs, e.g., traffic load, spectrum usage of BSs and the battery
capacity of MUs, is equivalent to or can be transformed to the nearest
BS cell association scheme.
\end{rem}

\begin{rem}
For log-normal shadowing, the condition of $\mathbb{E}_{\mathcal{H}_{n}^{\textrm{U}}}\left[\left(\mathcal{H}_{n}^{\textrm{U}}\right)^{2/\alpha_{n}^{\textrm{U}}}\right]<\infty$
is satisfied. While for a general case of shadowing or multi-path
fading model, $\mathbb{E}_{\mathcal{H}_{n}^{\textrm{U}}}\left[\left(\mathcal{H}_{n}^{\textrm{U}}\right)^{2/\alpha_{n}^{\textrm{U}}}\right]<\infty$
can also be easily met due to the bounded fading in practice. 
\end{rem}
In the next subsection, we will provide an application of the Equivalence
theorem, i.e., using the equivalence theorem to derive the distribution
of the strongest received signal power.

\subsection{The Distribution of the Strongest Received Signal Power}

In this subsection, we use stochastic geometry and Theorem \ref{thm: Equivalence theorem}
to obtain the distribution of the strongest received signal power.
Then we will use simulation results to validate our theoretical analysis.
\begin{lem}
\label{lem:CDF_strongest}Denote the strongest received signal power
as $\mathcal{P}$, i.e., $\mathcal{P}=\max\left(P_{i}\right)$, the
distribution of the strongest received signal power by the typical
MU can be given by
\begin{equation}
\Pr\left[\mathcal{P}\leqslant\gamma\right]=\exp\left[-\Lambda^{\textrm{NL}}\left(\left[0,\gamma^{-1/\alpha^{\textrm{NL}}}\right]\right)-\Lambda^{\textrm{L}}\left(\left[0,\gamma^{-1/\alpha^{\textrm{L}}}\right]\right)\right],\label{eq:CDF_power}
\end{equation}
where $\Lambda^{\textrm{NL}}\left(\left[0,t\right]\right)$ and $\Lambda^{\textrm{L}}\left(\left[0,t\right]\right)$
are defined in Eq. (\ref{eq:Measure_N}) and Eq. (\ref{eq:Measure_L}),
respectively.
\end{lem}
\begin{IEEEproof}
See Appendix B.
\end{IEEEproof}
If a specific NLoS/LoS transmission model is given, the distribution
of the strongest received signal power can be easily derived using
Lemma \ref{lem:CDF_strongest}. The following is an example assuming
that the LoS transmission probability follows a negative exponential
distribution.

\begin{figure}
\begin{centering}
\includegraphics[width=9cm]{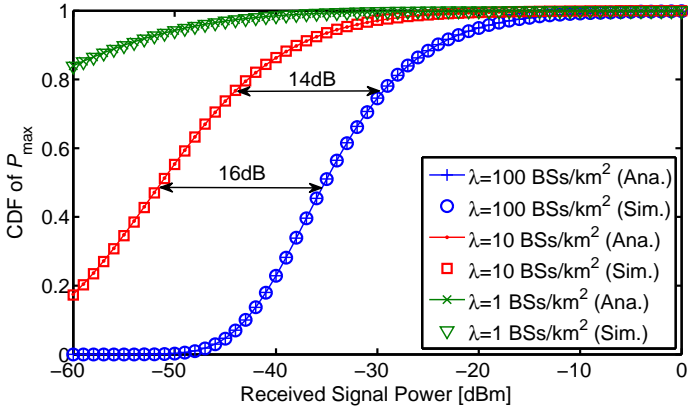}
\par\end{centering}
\caption{\label{fig:CDF-of-strongest}CDF of the strongest received signal
power, $P_{t}=1$ W (30 dBm), log-normal shadowing with zero means,
$\sigma^{\textrm{NL}}=4$ dB and $\sigma^{\textrm{L}}=3$ dB, simulation
and analytical results.}
\end{figure}

Let's consider a special case which assumes that $N=2$, $l_{1}^{\textrm{NL}}\left(R_{i}\right)=l_{2}^{\textrm{NL}}\left(R_{i}\right)=B^{\textrm{NL}}\left(R_{i}\right)^{-\alpha^{\textrm{NL}}}$,
$l_{1}^{\textrm{L}}\left(R_{i}\right)=l_{2}^{\textrm{L}}\left(R_{i}\right)=B^{\textrm{L}}\left(R_{i}\right)^{-\alpha^{\textrm{L}}}$
and $p_{1}^{\textrm{L}}\left(R_{i}\right)=p_{2}^{\textrm{L}}\left(R_{i}\right)=e^{-\kappa R_{i}}$,
where $\kappa$ is a constant determined by the density and the mean
length of blockages lying in the visual path between the typical MU
and the connected BS \cite{Bai15Coverage}, then the CDF of the strongest
received signal power is given by Eq. (\ref{eq:CDF_power}). Fig.
\ref{fig:CDF-of-strongest} illustrates the CDF of the strongest received
signal power and it can be seen that the simulation results perfectly
match the analytical results. From Fig. \ref{fig:CDF-of-strongest},
we can find that over 50\% of the strongest received signal power
is larger than -51 dBm when $\lambda=10\textrm{ BSs/k\ensuremath{m^{2}}}$
and this value increases by approximately 16 dB when $\lambda=10\textrm{0 BSs/k\ensuremath{m^{2}}}$,
which indicates that the strongest received signal power improves
as the BS density increases.

\section{\label{sec:SINR-Coverage-Probability}The Coverage Probability and
ASE Analysis}

In downlink performance evaluation, for networks where BSs are random
distributed according to a homogeneous PPP, it is sufficient to study
the performance of the typical MU located at the origin $o$ to characterize
the performance of a SCN using the Palm theory \cite[Eq. (4.71)]{Chiu13Stochastic}.
In this section, the coverage probability and ASE are first investigated
and then several special cases will be studied. 

\subsection{General Case and Main Result}

The coverage probability is generally defined as the probability that
the typical MU's measured SINR is greater than a designated threshold
$T$, i.e.,
\begin{equation}
p_{c}\left(\lambda,T\right)=\Pr\left[\textrm{SINR}>T\right],
\end{equation}
where the definition of SINR is given by Eq. (\ref{eq:SINR}) and
the subscript $i$ is omitted here for simplicity. Now, we present
a main result in this section on the coverage probability as follows.
\begin{thm}[Coverage Probability]
\label{thm:Pcoverage } Given that the signal propagation model follows
Eq. (\ref{eq:P_received}) and the typical MU selects the serving
BS according to Eq. (\ref{eq:inspower}) or Eq. (\ref{eq:avepower}),
then the coverage probability $p_{c}\left(\lambda,T\right)$ can be
evaluated by
\begin{equation}
p_{c}\left(\lambda,T\right)=\stackrel[n=1]{N}{\sum}p_{c,n}^{\textrm{L}}\left(\lambda,T\right)+\stackrel[n=1]{N}{\sum}p_{c,n}^{\textrm{NL}}\left(\lambda,T\right),\label{eq:theorem_pc}
\end{equation}
where
\begin{align}
 & p_{c,n}^{\textrm{L}}\left(\lambda,T\right)=\int_{y=0}^{\infty}\int_{\omega=-\infty}^{\infty}\left[\frac{1-e^{-j\omega/T}}{2\pi j\omega}\right]\lambda_{n}^{\textrm{L}}\left(y\right)\nonumber \\
 & \quad\,\times\exp\biggl\{-\Lambda_{n}^{\textrm{NL}}\left(\left[0,y^{\alpha_{n}^{\textrm{L}}/\alpha_{n}^{\textrm{NL}}}\right]\right)-\Lambda_{n}^{\textrm{L}}\left(\left[0,y\right]\right)+j\omega\eta y^{\alpha_{n}^{\textrm{L}}}\nonumber \\
 & \quad\,+\int_{t=y^{\alpha_{n}^{\textrm{L}}/\alpha_{n}^{\textrm{NL}}}}^{\infty}\left[e^{j\omega y^{\alpha_{n}^{\textrm{L}}}t^{-\alpha_{n}^{\textrm{NL}}}}-1\right]\lambda_{n}^{\textrm{NL}}\left(t\right)\textrm{d}t\nonumber \\
 & \quad\,+\int_{t=y}^{\infty}\left[e^{j\omega\left(y/t\right)^{\alpha_{n}^{\textrm{L}}}}-1\right]\lambda_{n}^{\textrm{L}}\left(t\right)\textrm{d}t\biggr\}\textrm{d}\omega\textrm{d}y\label{eq:theorem_pcL}
\end{align}
and
\begin{align}
 & p_{c,n}^{\textrm{NL}}\left(\lambda,T\right)=\int_{y=0}^{\infty}\int_{\omega=-\infty}^{\infty}\left[\frac{1-e^{-j\omega/T}}{2\pi j\omega}\right]\lambda_{n}^{\textrm{NL}}\left(y\right)\nonumber \\
 & \quad\,\times\exp\biggl\{-\Lambda_{n}^{\textrm{L}}\left(\left[0,y^{\alpha_{n}^{\textrm{NL}}/\alpha_{n}^{\textrm{L}}}\right]\right)-\Lambda_{n}^{\textrm{NL}}\left(\left[0,y\right]\right)+j\omega\eta y^{\alpha_{n}^{\textrm{NL}}}\nonumber \\
 & \quad\,+\int_{t=y^{\alpha_{n}^{\textrm{NL}}/\alpha_{n}^{\textrm{L}}}}^{\infty}\left[e^{j\omega y^{\alpha_{n}^{\textrm{NL}}}t^{-\alpha_{n}^{\textrm{L}}}}-1\right]\lambda_{n}^{\textrm{L}}\left(t\right)\textrm{d}t\nonumber \\
 & \quad\,+\int_{t=y}^{\infty}\left[e^{j\omega\left(y/t\right)^{\alpha_{n}^{\textrm{NL}}}}-1\right]\lambda_{n}^{\textrm{NL}}\left(t\right)\textrm{d}t\biggr\}\textrm{d}\omega\textrm{d}y,\label{eq:theorem_pcN}
\end{align}
where $j=\sqrt{-1}$ denotes the imaginary unit, $\lambda_{n}^{\textrm{NL}}\left(\cdot\right)$
and $\lambda_{n}^{\textrm{L}}\left(\cdot\right)$ are defined in Theorem
\ref{thm: Equivalence theorem}.
\end{thm}
\begin{IEEEproof}
See Appendix C.
\end{IEEEproof}
The coverage probability evaluated by Eq. (\ref{eq:theorem_pc}) in
Theorem \ref{thm:Pcoverage } is at least a 3-fold integral which
is somehow complicated for numerical computation. However, Theorem
\ref{thm:Pcoverage } gives general results that can be applied to
various multi-path fading or shadowing models, e.g., Rayleigh fading,
Nakagami-$m$ fading, etc, and various NLoS/LoS transmission models
as well. In the following, we turn our attention to a few relevant
special cases where 
\begin{enumerate}
\item NLoS transmissions and LoS transmissions are concatenated with different
shadowing, which will be studied in Subsection \ref{subsec:NLOS-transmissions-and};
\item NLoS transmissions and LoS transmissions are concatenated with Nakagami-$m$
fading of different parameters, which will be studied in Subsection
\ref{subsec:NLOS-and-LOS};
\item NLoS transmissions and LoS transmissions are concatenated with Rayleigh
fading and Rician fading, respectively, which will be studied in Subsection
\ref{subsec:NLOS-Transmission-+}.
\item Composite Rayleigh fading, Rician fading and log-normal shadowing
are considered in Subsection \ref{subsec:Composite-Rayleigh-Fading}.
\end{enumerate}

\subsection{\label{subsec:NLOS-transmissions-and}NLoS transmissions and LoS
transmissions are concatenated with different shadowing}

In the subsection, we assume that NLoS transmission and LoS transmission
are concatenated with different log-normal shadowing. The association
scheme is based on the SARP. Moreover, a simplified NLoS/LoS transmission
model is used for a specific analysis, which is expressed by
\begin{equation}
p^{\textrm{L}}\left(R_{i}\right)=\begin{cases}
1, & \hspace{-0.3cm}R_{i}\in\left(0,d\right]\\
0, & \hspace{-0.3cm}R_{i}\in\left(d,\infty\right]
\end{cases},
\end{equation}
where $d$ is a constant distance below which all BSs connect with
the typical MU with LoS transmissions. This model has been used in
some recent work \cite{Bai15Coverage,Singh15Tractable}. With assumptions
above, the intensity measure for NLoS transmissions, i.e., $\Lambda_{\log}^{\textrm{NL}}\left(\cdot\right)$,
is expressed as follows
\begin{align}
 & \quad\,\Lambda_{\log}^{\textrm{NL}}\left(\left[0,t\right]\right)=\mathbb{E}_{\mathcal{H}^{\textrm{NL}}}\left[2\pi\lambda\int_{R_{i}=0}^{t\left(B^{\textrm{NL}}\mathcal{H}^{\textrm{NL}}\right)^{1/\alpha^{\textrm{NL}}}}p^{\textrm{NL}}\left(R_{i}\right)R_{i}\textrm{d}R_{i}\right]\nonumber \\
 & =\frac{1}{2}\pi\lambda t^{2}\left(B^{\textrm{NL}}\right)^{2/\alpha^{\textrm{NL}}}e^{1/M_{\textrm{NL}}^{2}}\textrm{erfc}\left[M_{\textrm{NL}}\ln t+Q_{\textrm{NL}}\right]\nonumber \\
 & \quad\,-\frac{1}{2}\pi\lambda d^{2}\textrm{erfc}\left[M_{\textrm{NL}}\ln t+V_{\textrm{NL}}\right],\label{eq:Measure_N_lonormal}
\end{align}
where $\textrm{erfc}\left(\cdot\right)$ is the complementary error
function, $M_{\textrm{NL}}=-\frac{\alpha^{\textrm{NL}}}{\sqrt{2}\sigma^{\textrm{NL}}}$,
$Q_{\textrm{NL}}=\frac{\alpha^{\textrm{NL}}\ln d-\ln B^{\textrm{NL}}}{\sqrt{2}\sigma^{\textrm{NL}}}-\frac{1}{M_{\textrm{NL}}}$
and $V_{\textrm{NL}}=\frac{\alpha^{\textrm{NL}}\ln d-\ln B^{\textrm{NL}}}{\sqrt{2}\sigma^{\textrm{NL}}}$
are all constants. After obtaining $\Lambda_{\log}^{\textrm{NL}}\left(\cdot\right)$,
the density of NLoS BSs, i.e., $\lambda_{\log}^{\textrm{NL}}\left(\cdot\right)$,
can be readily derived as follows
\begin{align}
 & \quad\,\lambda_{\log}^{\textrm{NL}}\left(t\right)=\frac{\textrm{d}}{\textrm{d}t}\Lambda^{\textrm{NL}}\left(\left[0,t\right]\right)\nonumber \\
 & =\pi\lambda t\left(B^{\textrm{NL}}\right)^{2/\alpha^{\textrm{NL}}}e^{1/M_{\textrm{NL}}^{2}}\textrm{erfc}\left[M_{\textrm{NL}}\ln t+Q_{\textrm{NL}}\right]\nonumber \\
 & \quad\,+\frac{M_{\textrm{NL}}\lambda\sqrt{\pi}d^{2}}{t}e^{-\left(M_{\textrm{NL}}\ln t+V_{\textrm{NL}}\right)^{2}}\nonumber \\
 & \quad\,-M_{\textrm{NL}}\lambda t\sqrt{\pi}\left(B^{\textrm{NL}}\right)^{2/\alpha^{\textrm{NL}}}e^{1/M_{\textrm{NL}}^{2}-\left(M_{\textrm{NL}}\ln t+Q_{\textrm{NL}}\right)^{2}}.\label{eq:Intensity_N_lonormal}
\end{align}
Similarly, the intensity measure and density for LoS BSs are
\begin{align}
\Lambda_{\log}^{\textrm{L}}\left(\left[0,t\right]\right) & =\frac{1}{2}\pi\lambda t^{2}\left(B^{\textrm{L}}\right)^{2/\alpha^{\textrm{L}}}e^{1/M_{\textrm{L}}^{2}}\textrm{erfc}\left[M_{\textrm{L}}\ln t+Q_{\textrm{L}}\right]\nonumber \\
 & \quad\,+\frac{1}{2}\pi\lambda d^{2}\textrm{erfc}\left[-M_{\textrm{L}}\ln t+V_{\textrm{L}}\right],\label{eq:Measure_L_lonormal}
\end{align}

\begin{align}
\lambda_{\log}^{\textrm{L}}\left(t\right) & =\pi\lambda t\left(B^{\textrm{L}}\right)^{2/\alpha^{\textrm{L}}}e^{1/M_{\textrm{L}}^{2}}\textrm{erfc}\left[M_{\textrm{L}}\ln t+Q_{\textrm{L}}\right]\nonumber \\
 & \quad\,+\frac{M_{\textrm{L}}\lambda\sqrt{\pi}d^{2}}{t}e^{-\left(-M_{\textrm{L}}\ln t+V_{\textrm{L}}\right)^{2}}\nonumber \\
 & \quad\,-M_{\textrm{L}}\lambda t\sqrt{\pi}\left(B^{\textrm{L}}\right)^{2/\alpha^{\textrm{L}}}e^{1/M_{\textrm{L}}^{2}-\left(M_{\textrm{L}}\ln t+Q_{\textrm{L}}\right)^{2}},\label{eq:Intensity_L_lonormal}
\end{align}
respectively, where $M_{\textrm{L}}=\frac{\alpha^{\textrm{L}}}{\sqrt{2}\sigma^{\textrm{L}}}$,
$Q_{\textrm{L}}=\frac{\ln B^{\textrm{L}}-\alpha^{\textrm{L}}\ln d}{\sqrt{2}\sigma^{\textrm{L}}}+\frac{1}{M_{\textrm{L}}}$
and $V_{\textrm{L}}=\frac{\alpha^{\textrm{L}}\ln d-\ln B^{\textrm{L}}}{\sqrt{2}\sigma^{\textrm{L}}}$
are all constants. By substituting $\lambda_{\log}^{\textrm{NL}}\left(\cdot\right)$
and $\lambda_{\log}^{\textrm{L}}\left(\cdot\right)$ above into Eq.
(\ref{eq:theorem_pcL}) and Eq. (\ref{eq:theorem_pcN}), the coverage
probability can be obtained in this specific scenario, followed by
results in Section \ref{sec:Simulations}.

In the above scenario, the shadowing follows log-normal distributions.
However, Theorem \ref{thm:Pcoverage } can also be applied to a generalized
fading model and the coverage probability will be derived in the next
two sections.

\subsection{\label{subsec:NLOS-and-LOS}NLoS and LoS Transmissions are Concatenated
with Nakagami-$m$ Fading}

Note that if we replace $\mathcal{H}^{\textrm{U}}$ by multi-path
fading, i.e., $h^{\textrm{U}}$, Theorem \ref{thm:Pcoverage } also
works for the scenario where the SIRP association is applied. In this
subsection, we assume that both NLoS and LoS transmissions are concatenated
with Nakagami-$m$ fading of different parameters, e.g., $m^{\textrm{NL}}$
and $m^{\textrm{L}}$, then the channel power gains are distributed
according to Gamma distributions. That is, 
\begin{equation}
f_{h^{\textrm{U}}}\left(h\right)=\frac{\left(m^{\textrm{U}}\right)^{m^{\textrm{U}}}}{\Gamma\left(m^{\textrm{U}}\right)}h^{m^{\textrm{U}}-1}e^{-m^{\textrm{U}}h}.
\end{equation}

By substituting the PDF of $h^{\textrm{U}}$ into Eq. (\ref{eq:Lambda_N})
\textendash{} Eq. (\ref{eq:Measure_L}), the intensity measures and
intensities of $\overline{\Phi^{\textrm{NL}}}$ and $\overline{\Phi^{\textrm{L}}}$
can be readily obtained as follows

\begin{align}
 & \Lambda_{\textrm{Naka}}^{\textrm{NL}}\left(\left[0,t\right]\right)=-\frac{\pi\lambda d^{2}}{\Gamma\left(m^{\textrm{NL}}\right)}\Gamma\left(m^{\textrm{NL}},\frac{m^{\textrm{NL}}}{B^{\textrm{NL}}}\left(\frac{d}{t}\right)^{\alpha^{\textrm{NL}}}\right)\nonumber \\
 & +\frac{\pi\lambda t^{2}}{\Gamma\left(m^{\textrm{NL}}\right)}\left(\frac{B^{\textrm{NL}}}{m^{\textrm{NL}}}\right)^{\frac{2}{\alpha^{\textrm{NL}}}}\Gamma\left(\frac{2}{\alpha^{\textrm{NL}}}+m^{\textrm{NL}},\frac{m^{\textrm{NL}}}{B^{\textrm{NL}}}\left(\frac{d}{t}\right)^{\alpha^{\textrm{NL}}}\right),\label{eq:Measure_N_Nakagami}
\end{align}
\begin{align}
 & \Lambda_{\textrm{Naka}}^{\textrm{L}}\left(\left[0,t\right]\right)=\frac{\pi\lambda d^{2}}{\Gamma\left(m^{\textrm{L}}\right)}\Gamma\left(m^{\textrm{L}},\frac{m^{\textrm{L}}}{B^{\textrm{L}}}\left(\frac{d}{t}\right)^{\alpha^{\textrm{L}}}\right)\nonumber \\
 & +\frac{\pi\lambda t^{2}}{\Gamma\left(m^{\textrm{L}}\right)}\left(\frac{B^{\textrm{L}}}{m^{\textrm{L}}}\right)^{\frac{2}{\alpha^{\textrm{L}}}}\gamma\left(\frac{2}{\alpha^{\textrm{L}}}+m^{\textrm{L}},\frac{m^{\textrm{L}}}{B^{\textrm{L}}}\left(\frac{d}{t}\right)^{\alpha^{\textrm{L}}}\right),\label{eq:Measure_L_Nakagami}
\end{align}
\begin{align}
\lambda_{\textrm{Naka}}^{\textrm{NL}}\left(t\right) & =\frac{2\pi\lambda t}{\Gamma\left(m^{\textrm{NL}}\right)}\left(\frac{B^{\textrm{NL}}}{m^{\textrm{NL}}}\right)^{\frac{2}{\alpha^{\textrm{NL}}}}\nonumber \\
 & \quad\,\times\Gamma\left(\frac{2}{\alpha^{\textrm{NL}}}+m^{\textrm{NL}},\frac{m^{\textrm{NL}}}{B^{\textrm{NL}}}\left(\frac{d}{t}\right)^{\alpha^{\textrm{NL}}}\right),\label{eq:Intensity_N_Nakagami}
\end{align}
and
\begin{equation}
\lambda_{\textrm{Naka}}^{\textrm{L}}\left(t\right)=\frac{2\pi\lambda t}{\Gamma\left(m^{\textrm{L}}\right)}\left(\frac{B^{\textrm{L}}}{m^{\textrm{L}}}\right)^{\frac{2}{\alpha^{\textrm{L}}}}\gamma\left(\frac{2}{\alpha^{\textrm{L}}}+m^{\textrm{L}},\frac{m^{\textrm{L}}}{B^{\textrm{L}}}\left(\frac{d}{t}\right)^{\alpha^{\textrm{L}}}\right),\label{eq:Intensity_L_Nakagami}
\end{equation}
respectively, where $\Gamma\left(s,x\right)=\int_{x}^{\infty}v^{s-1}e^{-v}\textrm{d}v$
and $\gamma\left(s,x\right)=\int_{0}^{x}v^{s-1}e^{-v}\textrm{d}v$
denote the upper and the lower incomplete gamma functions, respectively,
$\Gamma\left(s\right)=\int_{0}^{\infty}v^{s-1}e^{-v}\textrm{d}v$
is the gamma function. The intermediate steps are easy to derive and
thus omitted here. By incorporating Eq. (\ref{eq:Measure_N_Nakagami})
- (\ref{eq:Intensity_L_Nakagami}) into Eq. (\ref{eq:theorem_pcL})
and Eq. (\ref{eq:theorem_pcN}), the coverage probability of a SCN
experiencing Nakagami-$m$ fading can be calculated.

\subsection{\label{subsec:NLOS-Transmission-+}NLoS Transmission + Rayleigh Fading
and LoS Transmission + Rician Fading}

In this part, we consider a more common case in which NLoS transmission
and LoS transmission are concatenated with Rayleigh fading and Rician
fading, respectively, i.e., $h^{\textrm{NL}}$ follows an exponential
distribution and $h^{\textrm{L}}$ follows a non-central Chi-squared
distribution. With $m=\left(K+1\right)^{2}/2K+1$, Rician fading can
be approximated by a Nakagami-$m$ distribution~\cite{Goldsmith05Wireless},
where $K$ is the Rician $K$-factor representing the ratio between
the power of the direct path and that of the scattered paths. Without
loss of generality, we assume $f_{h^{\textrm{NL}}}\left(h\right)=e^{-h}$
and $f_{h^{\textrm{L}}}\left(h\right)=\frac{m^{m}}{\Gamma\left(m\right)}h^{m-1}e^{-mh}$
for NLoS and LoS transmissions, respectively.

As we have provided the intensity measure and intensity of $\overline{\Phi^{\textrm{L}}}$
experiencing Nakagami-$m$ fading in the previous subsection, in this
part we just provide the intensity measures and intensities of $\overline{\Phi^{\textrm{NL}}}$.
By substituting the PDF of $h^{\textrm{NL}}$ into Eq. (\ref{eq:Measure_N})
and Eq. (\ref{eq:Lambda_N}), $\Lambda_{\textrm{Ray}}^{\textrm{NL}}\left(\left[0,t\right]\right)$
and $\lambda_{\textrm{Ray}}^{\textrm{NL}}\left(t\right)$ can be easily
evaluated by
\begin{align}
\Lambda_{\textrm{Ray}}^{\textrm{NL}}\left(\left[0,t\right]\right) & =\pi\lambda t^{2}\left(B^{\textrm{NL}}\right)^{\frac{2}{\alpha^{\textrm{NL}}}}\Gamma\left(\frac{2}{\alpha^{\textrm{NL}}}+1,\frac{1}{B^{\textrm{NL}}}\left(\frac{d}{t}\right)^{\alpha^{\textrm{NL}}}\right)\nonumber \\
 & -\pi\lambda d^{2}\exp\left[-\frac{\left(d/t\right)^{\alpha^{\textrm{NL}}}}{B^{\textrm{NL}}}\right],\label{eq:Measure_N_Rayleigh}
\end{align}
and
\begin{equation}
\lambda_{\textrm{Ray}}^{\textrm{NL}}\left(t\right)=2\pi\lambda t\left(B^{\textrm{NL}}\right)^{\frac{2}{\alpha^{\textrm{NL}}}}\Gamma\left(\frac{2}{\alpha^{\textrm{NL}}}+1,\frac{1}{B^{\textrm{NL}}}\left(\frac{d}{t}\right)^{\alpha^{\textrm{NL}}}\right),\label{eq:Intensity_N_Rayleigh}
\end{equation}
respectively. After substituting the intensity measures and intensities
of $\overline{\Phi^{\textrm{NL}}}$ and $\overline{\Phi^{\textrm{L}}}$
into Eq. (\ref{eq:theorem_pcN}) and Eq. (\ref{eq:theorem_pcL}),
the coverage probability can be obtained and we omit the rest derivations.

\subsection{\label{subsec:Composite-Rayleigh-Fading}Composite Rayleigh Fading,
Rician Fading and Log-normal Shadowing}

Inspired by \cite{Renzo13Average} which takes composite fading into
consideration, in this subsection both fading and shadowing will be
considered simultaneously. In \cite{Liu16Optimal}, a channel gain
PDF which characterizes the composite effect of Rayleigh fading and
log-normal shadowing is given by
\begin{equation}
f_{H}\left(h\right)=\frac{1}{\sqrt{2\pi\sigma_{s}^{2}}}\int_{x=0}^{\infty}\frac{1}{x^{2}}e^{-\frac{h}{x}-\frac{\left(\ln x-\mu_{s}\right)^{2}}{2\sigma_{s}^{2}}}\textrm{d}x,\label{eq:pdf_composite_Rayleigh_lognormal}
\end{equation}
where $\mu_{s}$ and $\sigma_{s}^{2}$ are the mean and variance of
log-normal shadowing, respectively. By substituting the PDF of $H$
into Eq. (\ref{eq:Measure_N}) and Eq. (\ref{eq:Lambda_N}), $\Lambda^{\textrm{NL}}\left(\left[0,t\right]\right)$
and $\lambda^{\textrm{NL}}\left(t\right)$ can be obtained, which
however are non-closed forms. As for the channel model with composite
Rician fading and log-normal shadowing, no such PDF could be found
like Eq. (\ref{eq:pdf_composite_Rayleigh_lognormal}). In this context,
we utilize a simplified composite fading and shadowing channel model
in which the desired signal experiences Rayleigh fading or Rician
fading and the interference signal experiences log-normal shadowing
\cite{Andrews11A,Ge15Spatial}. For example, assume that the desired
NLoS transmission is concatenated with Rayleigh fading, the desired
LoS transmission is concatenated with Rician fading and the aggregate
interference is concatenated with log-normal shadowing. The coverage
probability can be readily obtained by substituting $\lambda_{\log}^{\textrm{NL}}\left(t\right)$
into Eq. (\ref{eq:proof_E_N}), $\lambda_{\log}^{\textrm{NL}}\left(t\right)$
into Eq. (\ref{eq:proof_E_L}) and $\lambda_{\textrm{Ray}}^{\textrm{U}}\left(t\right)$,
$\Lambda_{\textrm{Ray}}^{\textrm{U}}\left(\left[0,t\right]\right)$
into Eq. (\ref{eq:pdf_y_N}), respectively.

\subsection{The Asymptotic Analysis}

In the following, an asymptotic analysis will be given for the situation
where BS deployment becomes ultra-dense, i.e., $\lambda\rightarrow\infty$,
which helps to analyze the performance with a concise form.
\begin{cor}
\label{cor: Pc_asymptotic}If $T\geqslant1$, the coverage probability
of $p_{c}\left(\lambda,T\right)$ considering a single-slope path
loss model in Eq. (\ref{eq:theorem_pc}) when $\lambda\rightarrow\infty$
converges as follows
\begin{align}
\underset{\lambda\rightarrow\infty}{\lim}p_{c}\left(\lambda,T\right) & =\underset{\lambda\rightarrow\infty}{\lim}\Pr\left[\textrm{SINR}>T\right]\nonumber \\
 & \overset{\left(a\right)}{=}\underset{\lambda\rightarrow\infty}{\lim}\Pr\left[\textrm{SIR}>T\right]\nonumber \\
 & \overset{\left(b\right)}{=}\frac{\alpha^{\textrm{L}}\sin\left(2\pi/\alpha^{\textrm{L}}\right)}{2\pi T^{2/\alpha^{\textrm{L}}}}.\label{eq:Pc corollary}
\end{align}
\end{cor}
\begin{IEEEproof}
A sketch of the proof of Corollary \ref{cor: Pc_asymptotic} is given
here. In Eq. (\ref{eq:Pc corollary}), $\left(a\right)$ is due to
the reason that when $\lambda\rightarrow\infty$, the network is interference-limited
and noise can be ignored compared with the aggregate interference,
which is also validated by results in Section \ref{sec:Simulations}.
The proof of $\left(b\right)$ can be found in \cite[Remark 9]{Blaszczyszyn13Using}
and \cite[Theorem 4]{Bai15Coverage} and are omitted here.
\end{IEEEproof}
From Corollary \ref{cor: Pc_asymptotic}, it can be concluded that
for dense SCNs the coverage probability is invariant with respect
to BS density $\lambda$ and even the distribution of shadowing/fading.
However, when the BS density is not dense enough, the coverage probability
reveals an interesting performance, which will be fully studied in
Section \ref{sec:Simulations}.

Moreover, when considering a multi-slope path model, when $\lambda\rightarrow\infty$,
the noise power can be ignored compared with the interference and
the typical MU will connected to a LoS BS almost for sure due the
blockage probability model in Eq. (\ref{eq:PL_BS2UE}). In this context,
\begin{align}
\underset{\lambda\rightarrow\infty}{\lim}p_{c}\left(\lambda,T\right) & =\underset{\lambda\rightarrow\infty}{\lim}\Pr\left[\textrm{SIR}>T\right]\nonumber \\
 & =\underset{\lambda\rightarrow\infty}{\lim}\Pr\left[\textrm{SIR}\left(\left\{ l_{n},\alpha_{n}^{\textrm{L}}\right\} \right)>T\right],
\end{align}
where $\Pr\left[\textrm{SIR}\left(\left\{ l_{n},\alpha_{n}^{\textrm{L}}\right\} \right)>T\right]$
denotes the coverage probability with multi-slope path loss model
($N$ piece-wise function) but only LoS transmissions being considered.
From \cite[Lemma 3]{Zhang15Downlink} and assuming that $0\leqslant\alpha_{1}^{\textrm{L}}\leqslant\alpha_{2}^{\textrm{L}}\leqslant\cdots\leqslant\alpha_{N}^{\textrm{L}}$,
when $\lambda\rightarrow\infty$, the coverage probability approaches
to
\begin{align}
\underset{\lambda\rightarrow\infty}{\lim}p_{c}\left(\lambda,T\right) & =\underset{\lambda\rightarrow\infty}{\lim}\Pr\left[\textrm{SIR}\left(\left\{ l_{n},\alpha_{n}^{\textrm{L}}\right\} \right)>T\right]\nonumber \\
 & =\underset{\lambda\rightarrow\infty}{\lim}\Pr\left[\textrm{SIR}\left(\left\{ l_{1},\alpha_{1}^{\textrm{L}}\right\} \right)>T\right],
\end{align}
which is only determined by the first piece single-slope path loss
function. As for the ASE scaling law against $\lambda$, the readers
may refer to \cite{Zhang15Downlink,Nguyen17Performance}.

\subsection{The ASE Upper Bound}

Finally, the upper bound of ASE in units of $\textrm{bps/Hz/k\ensuremath{m^{2}}}$
for a given BS density $\lambda$ can be derived as follows \cite{Ding16Performance}
\begin{align}
\textrm{ASE}\left(\lambda\right) & =\lambda\mathbb{E}_{\textrm{SINR}}\left[\log_{2}\left(1+\textrm{SINR}\right)\right]\nonumber \\
 & =\lambda\int_{u=T}^{\infty}\log_{2}\left(1+u\right)f_{\textrm{SINR}}\left(\lambda,u\right)\textrm{d}u\nonumber \\
 & \leqslant\lambda\int_{u=0}^{\infty}\log_{2}\left(1+u\right)f_{\textrm{SINR}}\left(\lambda,u\right)\textrm{d}u\nonumber \\
 & =\frac{\lambda}{\ln2}\int_{u=0}^{\infty}\frac{p_{c}\left(\lambda,T\right)}{u+1}\textrm{d}u,\label{eq:ASE}
\end{align}
where the integral in Eq. (\ref{eq:ASE}) can be numerically obtained
\cite[Eq. (10)]{Yilmaz12AUnified}. Note that in \cite{Renzo13Average},
the proposed MGF\textendash based approach can efficiently compute
the ASE instead of obtaining the coverage probability in advance.
While in our work, the coverage probability and the ASE can be analyzed
simultaneously at the expense of increased complexity of computation.

\section{\label{sec:Simulations}Simulations and Discussions}

This section presents numerical results to validate our analysis,
followed by discussions to shed new light on the performance of SCNs.
We use the following parameter values, $P_{t}=30\textrm{ dBm}$, $A^{\textrm{NL}}=30.8\textrm{ dB}$
, $A^{\textrm{L}}=2.7\textrm{ dB}$, $\alpha^{\textrm{NL}}=4.28$,
$\alpha^{\textrm{L}}=2.42$, $\sigma^{\textrm{NL}}=4\textrm{ dB}$,
$\sigma^{\textrm{L}}=3\textrm{ dB}$, $T=0\textrm{ dB}$ and $d=250\textrm{ m}$
\cite{Bai15Coverage,Singh15Tractable,3GPP36828,Yang16Coverage,Yang17PerformanceC,Ge13Modeling}.

\subsection{\label{subsec:Validation-of-the}Validation of the Analytical Results
of $p_{c}\left(\lambda,T\right)$ with Monte Carlo Simulations}

\begin{figure}
\begin{centering}
\includegraphics[width=9cm]{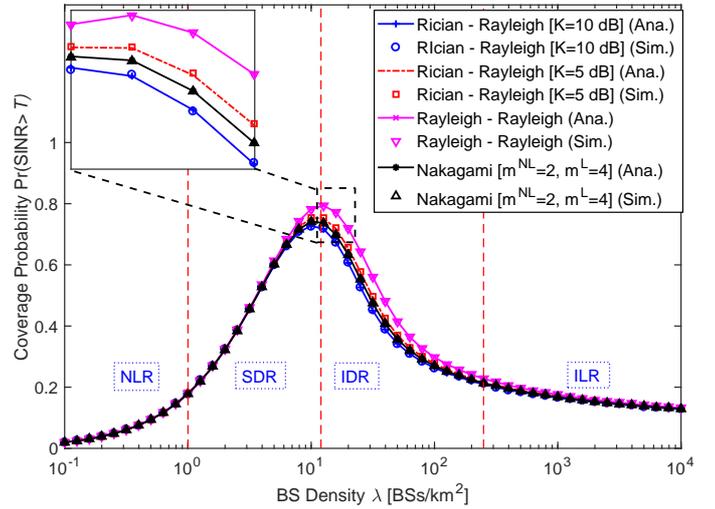}
\par\end{centering}
\caption{\label{fig:SINR-Coverage-Probability-density1}Coverage probability
vs. BS density $\lambda$, $\eta=-95\textrm{ dBm}$, $\mu^{\textrm{NL}}=\mu^{\textrm{L}}=1$,
simulation and analytical results.}
\end{figure}

\begin{figure}
\begin{centering}
\includegraphics[width=9cm]{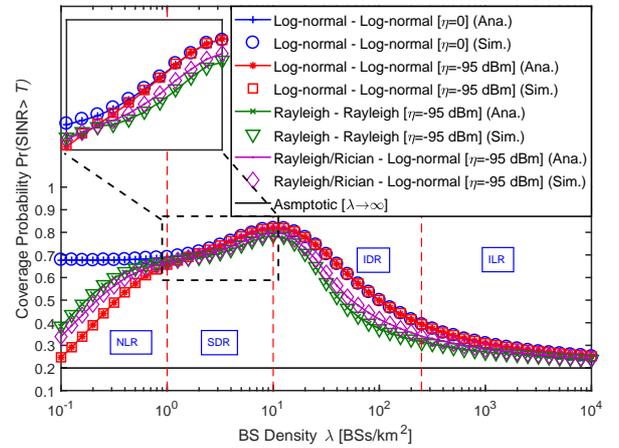}
\par\end{centering}
\caption{\label{fig:SINR-Coverage-Probability-density}Coverage probability
vs. BS density $\lambda$, $\mu^{\textrm{NL}}=\frac{1}{23.45}$, $\mu^{\textrm{L}}=\frac{1}{7.32}$
\cite{Singh15Tractable}, simulation and analytical results.}
\end{figure}

\begin{figure}
\begin{centering}
\includegraphics[width=9cm]{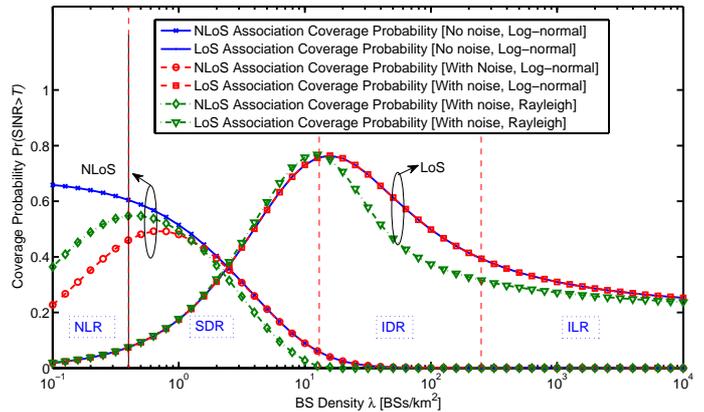}
\par\end{centering}
\caption{\label{fig:NLOS_LOS_Probability}NLoS/LoS association coverage probability
vs. BS density $\lambda$, $\mu^{\textrm{NL}}=\frac{1}{23.45}$, $\mu^{\textrm{L}}=\frac{1}{7.32}$
\cite{Singh15Tractable}.}
\end{figure}

\begin{figure}
\begin{centering}
\includegraphics[width=9cm]{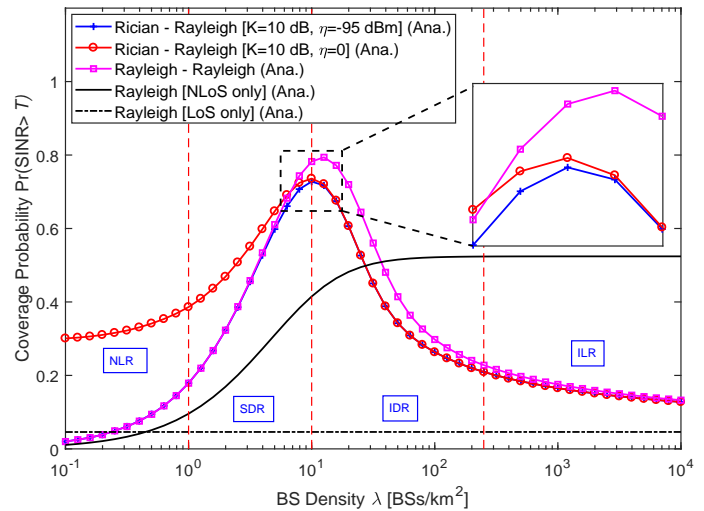}
\par\end{centering}
\caption{\label{fig:Comparison-of-different}Coverage probability vs. BS density
$\lambda$, $\mu^{\textrm{NL}}=\mu^{\textrm{L}}=1$.}
\end{figure}

The results of $p_{c}\left(\lambda,T\right)$ configured with $T=0\textrm{ dB}$
are plotted in Fig. \ref{fig:SINR-Coverage-Probability-density1}
and Fig. \ref{fig:SINR-Coverage-Probability-density}, which illustrate
the coverage performance of networks using SIRP and SARP, respectively.
As can be observed from Fig. \ref{fig:SINR-Coverage-Probability-density1}
and Fig. \ref{fig:SINR-Coverage-Probability-density}, the analytical
results match the simulation results well, which validate the accuracy
of our theoretical analysis. Note that in the case where both NLoS
and LoS transmissions are concatenated with Rayleigh fading, the coverage
probability is the highest among the interested cases. By contrast,
in the case where NLoS transmission is concatenated with Rayleigh
fading and LoS transmission is concatenated with Rician fading with
$K=10\textrm{ dB}$, the coverage probability is the lowest, which
suggests that Rayleigh fading model exaggerates network performance.
Meanwhile, we should notice that the gap between the plotted curves
is small, which means that multi-path fading has a minor impact on
the coverage probability performance. In Fig. \ref{fig:SINR-Coverage-Probability-density},
the coverage probability with composite fading and shadowing channel
model is also illustrated, which shows a similar tendency compared
with others. With the assistance of Fig. \ref{fig:NLOS_LOS_Probability},
we conclude that the performance of small cell networks can be divided
into four different regimes according to the density of small cell
BSs, where in each regime, the performance is dominated by different
factors. That is,

\begin{itemize}
\item \textbf{N}oise-\textbf{L}imited \textbf{R}egime (NLR): ($\lambda\leqslant1\textrm{ BSs/k\ensuremath{m^{2}}}$
in Fig. \ref{fig:SINR-Coverage-Probability-density}, Fig. \ref{fig:NLOS_LOS_Probability}
and Fig. \ref{fig:Comparison-of-different}). In this regime, the
typical MU is likely to have a NLoS path with the serving BS, see
Fig. \ref{fig:NLOS_LOS_Probability} . The network in the NLR regime
is very sparse and thus the interference can be ignored compared with
the thermal noise if we use \textbf{SINR} for performance metric.
In this case, $\textrm{SINR}=\frac{S}{\eta}$ and the coverage probability
will increase with the increase of $\lambda$ as the strongest received
power ($S$) will grow and noise power ($\eta$) will remain the same.
While if we use \textbf{SIR} for performance metric, the SIR coverage
probability remain almost stable in this regime as $\lambda$ increases.
This is because the increase in the received signal power is counterbalanced
by the increase in the aggregate interference power. Besides, as the
aggregate interference power is smaller than noise power, the SIR
coverage probability is larger than the SINR coverage probability.
\item \textbf{S}ignal\textbf{-D}ominated \textbf{R}egime (SDR): ($\lambda\in(1,10]\textrm{ BSs/k\ensuremath{m^{2}}}$
in Fig. \ref{fig:SINR-Coverage-Probability-density}, Fig. \ref{fig:NLOS_LOS_Probability}
and Fig. \ref{fig:Comparison-of-different}). In this regime, when
$\lambda$ is small, the typical MU has a higher probability to connect
to a NLoS BS; while when $\lambda$ becomes larger, the typical MU
has an increasingly higher probability to connect to a LoS BS. That
is to say, with the increase of $\lambda$, the typical MU is more
likely to be in LoS with the associated BS, i.e., the received signal
transforms from NLoS to LoS path. Even though the associated BS is
LoS, the majority of interfering BSs are still NLoS in this regime
and thus the SINR (or SIR) coverage probability keeps growing. From
this regime on, noise power has a negligible impact on coverage performance,
i.e., the SCN is interference-limited. Besides, if ignoring noise
power, from the NLR to the SDR, the coverage probability from NLoS
BSs decreases to almost zero and the coverage probability contributed
by LoS BSs increases. It is because when the network is sparse, almost
all MUs are associated with NLoS BSs and when the network goes denser,
MUs shift from NLoS BSs to LoS BSs.
\item \textbf{I}nterference\textbf{-D}ominated \textbf{R}egime (IDR): ($\lambda\in(10,250]\textrm{ BSs/k\ensuremath{m^{2}}}$
in Fig. \ref{fig:SINR-Coverage-Probability-density}, Fig. \ref{fig:NLOS_LOS_Probability}
and Fig. \ref{fig:Comparison-of-different}). In this regime, the
typical MU is connected to a LoS BS with a high probability. However,
different from the situation in the SDR, the majority of interfering
BSs experience transitions from NLoS to LoS path, which causes much
more severe interference to the typical MU compared with interfering
BSs with NLoS paths. As a result, the SINR (or SIR) coverage probability
decreases with the increase of $\lambda$ because the transition of
interference from NLoS path to LoS path causes a larger increase in
interference compared with that in signal. Note that in this regime
the coverage probability performance in our model exhibits a huge
difference from that of the analysis in \cite{Andrews11A}, which
are indicated as ``NLoS only'' and ``LoS only'' in Fig. \ref{fig:Comparison-of-different}.
\item \textbf{I}nterference-\textbf{L}imited \textbf{R}egime (ILR): ($\lambda>250\textrm{ BSs/k\ensuremath{m^{2}}}$
in Fig. \ref{fig:SINR-Coverage-Probability-density}, Fig. \ref{fig:NLOS_LOS_Probability}
and Fig. \ref{fig:Comparison-of-different}). In this regime, the
network is extremely dense and grow close to the LoS-BS-only scenario
as the increase of $\lambda$. The SINR (or SIR) coverage probability
will become stable with the increase in BS density as any increase
in the received LoS BS signal power is counterbalanced by the increase
in the aggregate LoS BS interference power, which is also illuminated
by Corollary \ref{cor: Pc_asymptotic}.
\end{itemize}

\begin{figure}
\begin{centering}
\includegraphics[width=9cm]{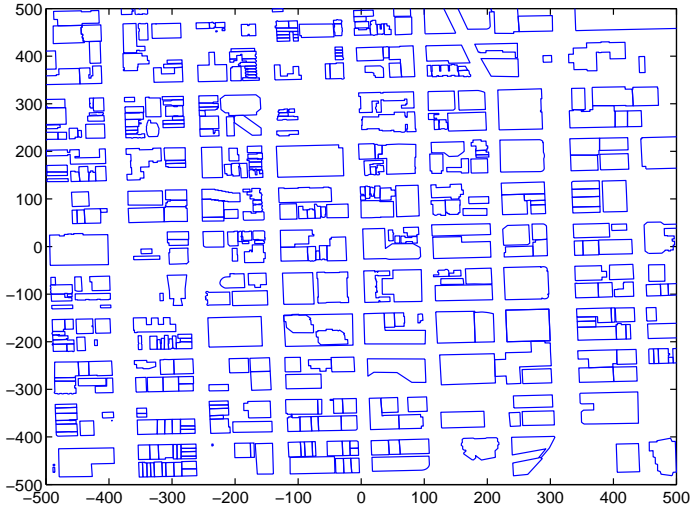}
\par\end{centering}
\caption{\label{fig:Building-topology}Building topology of Chicago.}
\end{figure}

\begin{figure}
\begin{centering}
\includegraphics[width=9cm]{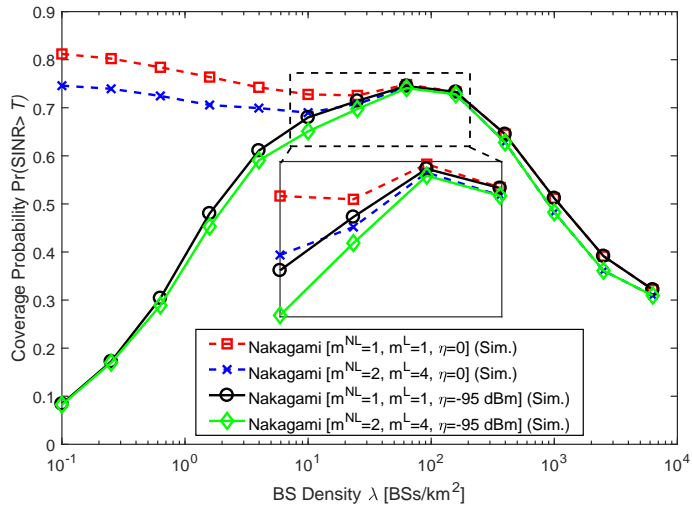}
\par\end{centering}
\caption{\label{fig:Pc Chicigo}Coverage probability vs. BS density in Chicago.}
\end{figure}

To validate the four performance regimes still exist in the networks
employing actual building topology. Followed by \cite{Singh15Tractable},
we present the coverage probability of Chicago in Fig. \ref{fig:Pc Chicigo}
whose topology is shown in Fig. \ref{fig:Building-topology}. Note
that the NLoS transmissions and LoS transmissions are not determined
by the one-parameter distance-based statistic model which is used
in our work. Instead, they are determined by whether the transmission
links are blocked by buildings or not. It is found that the four performance
regimes still exist especially when the noise power is considered
with a real building topology. The only difference is that the BS
density at which the coverage probability peaks shifts from around
$10\textrm{ BSs/k\ensuremath{m^{2}}}$ to around $100\textrm{ BSs/k\ensuremath{m^{2}}}$.
In our work, the probability function of blockage is a piece-wise
function which can be adjusted according to the real scenario.

\subsection{Boundary Definitions }

Based on the qualitative results above, it is interesting to develop
a qualitative definition of the boundaries among adjacent regimes.
In this subsection, we propose the following definition to characterize
three BS density boundaries, which makes the analysis of SCNS more
formal.
\begin{defn}
The boundary between NLR and SDR is $\lambda_{\textrm{SDR}}^{\textrm{NLR}}$
which is defined as follows
\begin{equation}
\lambda_{\textrm{SDR}}^{\textrm{NLR}}=\underset{\lambda}{\arg}\left\{ \mathbb{E}\left[I\right]=\eta\right\} .\label{eq:NLR-SDR}
\end{equation}
\end{defn}
The intuition of this definition is when $\lambda>\lambda_{\textrm{SN2LTR}}^{\textrm{NLR}}$,
the aggregate interference has a greater impact on network performance
than that caused by noise.
\begin{defn}
The boundary between SDR and IDR is $\lambda_{\textrm{IDR}}^{\textrm{SDR}}$,
which is defined as the BS density that the coverage probability achieves
the highest, i.e.,
\begin{equation}
p_{c}^{\max}=p_{c}\left(\lambda_{\textrm{IDR}}^{\textrm{SDR}},T\right),\label{eq:SDR-IDR}
\end{equation}
\end{defn}
which is equivalent to $\lambda_{\textrm{IDR}}^{\textrm{SDR}}=\underset{\lambda}{\arg\max}\left\{ p_{c}\left(\lambda,T\right)\right\} $.
The definition above reveals that $p_{c}\left(\lambda_{\textrm{IDR}}^{\textrm{SDR}},T\right)$
is the maximum coverage probability if other parameters are fixed.
From discussions above, the performance in the SDR is dominated by
the desired signal, while in the IDR, the performance is dominated
by the interference. When $\lambda>\lambda_{\textrm{IDR}}^{\textrm{SDR}}$,
LoS interference will degrade the coverage performance.
\begin{defn}
The boundary between IDR and ILR is $\lambda_{\textrm{ILR}}^{\textrm{IDR}}$,
which is defined as $\forall\lambda>\lambda_{\textrm{ILR}}^{\textrm{IDR}}$
\begin{equation}
\mathbb{E}\left[I\right]\gg\eta,\label{eq:IDR-ILR}
\end{equation}
\end{defn}
which is equivalent to $\lambda_{\textrm{ILR}}^{\textrm{IDR}}=\underset{\lambda}{\arg}\left\{ \mathbb{E}\left[I\right]=\epsilon\eta\right\} ,$
where $\epsilon\gg1$. When $\lambda$ becomes larger and larger,
the SCNs fall into the ILR, i.e., the aggregate interference might
be extremely large compared with the noise power $\eta$, which is
shown by Eq. (\ref{eq:IDR-ILR}). When $\lambda>\lambda_{\textrm{ILR}}^{\textrm{IDR}}$,
the coverage changes slowly and approaches the asymptotic value. In
the following, we will analyze the ASE performance in the four defined
regimes.

\subsection{\label{subsec:Discussion-on-the}Discussion on the Analytical Results
of the Upper Bound $\textrm{ASE}\left(\lambda\right)$}

\begin{figure}
\begin{centering}
\includegraphics[width=9cm]{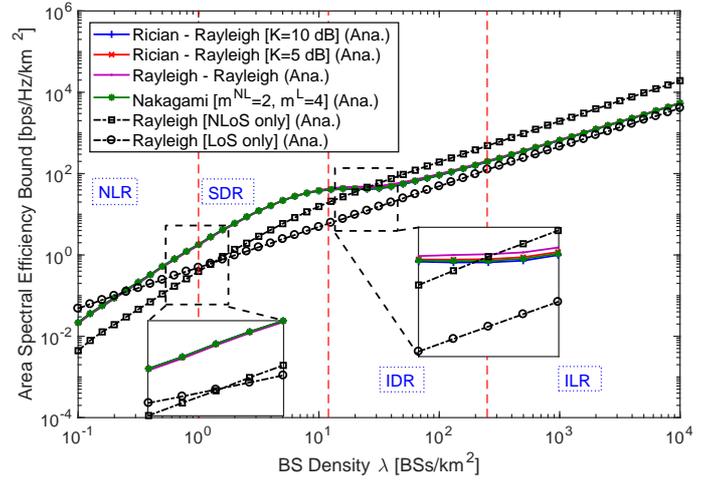}
\par\end{centering}
\caption{\label{fig:ASE-1}ASE vs. BS density $\lambda$, $\eta=-95\textrm{ dBm}$,
$\mu^{\textrm{NL}}=\mu^{\textrm{L}}=1$.}
\end{figure}

In this part, the upper bound of ASE with $T=0\textrm{ dB}$ is evaluated
analytically only, as the upper bound is a function of $p_{c}\left(\lambda,T\right)$
shown in Eq. (\ref{eq:ASE}).

Fig. \ref{fig:ASE-1} illustrates the upper bound with different fading
models vs. $\lambda$. It is found that the upper bound of the SCN
incorporating both NLoS and LoS transmissions reveal a deviation from
that of the analysis considering NLoS (or LoS) transmissions only
\cite{Andrews11A}. Specifically, when the SCN is sparse and thus
in the NLR or the SDR, the upper bound quickly increases with $\lambda$
because the network is generally noise-limited, and thus adding more
small cells immensely benefits the ASE. When the network becomes dense,
i.e., $\lambda$ enters the IDR, which is the practical range of $\lambda$
for the existing 4G networks and the future 5G networks, the trend
of the upper bound is very interesting. First, when $\lambda\in(10,50]\textrm{ BSs/k\ensuremath{m^{2}}}$,
the upper bound exhibits a slowing-down in the rate of growth due
to the fast decrease of the coverage probability at $\lambda\in(10,50]\textrm{ BSs/k\ensuremath{m^{2}}}$,
as shown in Fig. \ref{fig:SINR-Coverage-Probability-density1} and
Fig. \ref{fig:SINR-Coverage-Probability-density}. Second, when $\lambda>50\textrm{ BSs/k\ensuremath{m^{2}}}$,
the upper bound will pick up the growth rate since the decrease of
the coverage probability becomes a minor factor compared with the
increase of $\lambda$. When the SCN is extremely dense, e.g., $\lambda$
is in the ILR, the upper bound exhibits a nearly linear trajectory
with respect to $\lambda$ because both the signal power and the interference
power are now LoS dominated, and thus statistically stable as explained
before. Moreover, it can be observed that the change of the multi-path
fading model has a minor impact on the upper bound compared with the
change of the path loss model.

\subsection{Discussion on the Value of Theoretical Analysis }

Simulation is time consuming for $\lambda\geqslant10^{3}\textrm{ BSs/k\ensuremath{m^{2}}}$
and almost infeasible for $\lambda\geqslant10^{5}\textrm{ BSs/k\ensuremath{m^{2}}}$.
For example, simulation for networks with $\lambda=10^{5}\textrm{ BSs/k\ensuremath{m^{2}}}$
needs at least $4\times10^{5}$ BSs to get a smooth curve, which consumes
almost 2 weeks for a 8 core PC. On the other hand, the computational
complexity for theoretical analysis is stable for all BS densities.
In this context, the theoretical analysis is useful when you want
to analyze an ultra-dense network, i.e., $\lambda\geqslant10^{5}\textrm{ BSs/k\ensuremath{m^{2}}}$.

Based on the findings of NLoS-to-LoS-transition, next we will introduce
some guidance on how to design and manage the cellular networks in
order to optimize the network performance as we evolve into dense
SCNs. 

As described in section \ref{subsec:Validation-of-the} and \ref{subsec:Discussion-on-the},
the ASE increases almost for sure as SCNs becomes denser due to the
gain of frequency reuse. In contrast, the coverage probability of
SCNs will firstly increase and then decrease with the increase of
BS density $\lambda$. In this context, there is a trade off between
the coverage probability and the ASE in the future 5G SCNs incorporating
both NLoS and LoS transmissions. While in \cite{Andrews11A}, denser
SCNs always provide better network performance with respect to the
ASE as well as the coverage probability. It is noted that compared
with the existing work \cite{Andrews11A,Singh15Tractable} which assume
the network works either in the NLR or the ILR, our findings with
more elaborate working regimes partition provide guidance for network
design and optimization. A rough working regimes partition may not
give useful suggestions on network performance enhancement especially
in the transitional regimes between the NLR and the ILR. For example,
increasing BS transmit power can improve the coverage probability
in the NLR but fails in the ILR \cite{Andrews11A}. In the transitional
regimes between the NLR and the ILR, we may imagine that this technique
transforms form being useful to being useless. However, due to the
lack of detailed features in the transitional regimes, we are still
not sure whether to use this technique or not. Regarding the four
performance regimes, in the following, we try to provide different
techniques which can be used to enhance the network performance.
\begin{enumerate}
\item NLR: When the network works in the NLR, e.g., most mmWave network,
the interference is not the dominate factor and the desired signal
strength could be enhanced by utilizing BS power control, and directional
antennas, etc.
\item SDR: In this regime, the desired signal strength is still the dominate
factor. Thus the techniques used in the NLR to enhance the performance
are valid as well. However, some techniques will not as useful as
that in the NLR. For example, directional antenna technique may work
efficiently, but BS power control technique may be not so efficient
as in this regime increasing the transmit power may cause interference
to other users.
\item IDR: According to the data, the current 4G network is operating in
the SDR. As we deploy more and more BSs in the future to meet the
skyrocketing demands on wireless data, the network will fall into
the IDR. In this regime, we need elaborately design the network system
including transmission techniques, medium access control (MAC) protocols
and coding techniques, etc, to compensate the impair of network coverage
caused by strong LoS interference. The most common MAC protocols are
interference cancellation, interference avoidance, and interference
control. By jointly utilizing advanced transmission techniques like
beamforming techniques, multiple-input-multiple-output (MIMO), multi-antenna,
coordinated multi-point (CoMP) transmissions and better coding techniques,
the interference will be mitigated to an acceptable level, which benefits
both the coverage probability and the ASE a lot.
\item ILR: In this regime, the coverage performance is so poor even though
the ASE increases with the BS density. Techniques mentioned for IDR
should be already utilized in advance to avoid entering this regime.
\end{enumerate}

\section{\label{sec:Conclusions-and-Future}Conclusions and Future Work}

In this paper, we illustrated the transition behaviors in SCNs incorporating
both NLoS and LoS transmissions. Based on our analysis, the network
can be divided into four regimes, i.e., the NLR, the SDR, the IDR
and the ILR, where in each regime the performance is dominated by
different factors. The analysis helps to understand as the BS density
grows continually, which dominant factor that determines the cellular
network performance and therefore provide guidance on the design and
management of the cellular networks as we evolve into dense SCNs.
Moreover, our work adopt a generalized shadowing/fading model, in
which log-normal shadowing and/or Rayleigh fading can be treated in
a unified framework.

It is noted that constant transmit power is assumed in our work, however,
when the BS density, i.e., $\lambda$ is large, the BS transmit power
usually decreases to reduce the inter-cell interference. In our on-going
work, i.e., \cite{Ding17Performance,Yang17On}, density-dependent
transmit power is considered and the coverage probability and the
ASE reveal a different tendency. In our future work, shadowing and
multi-path fading model will be considered simultaneously which is
more practical for the real network. Furthermore, heterogeneous networks
(HetNets) incorporating both NLoS and LoS transmissions will also
be investigated. 

\section*{Appendix A: Proof of Theorem \ref{thm: Equivalence theorem}}

Firstly, we will obtain the intensity measure $\Lambda_{n}^{\textrm{NL}}$
of $\overline{\Phi_{n}^{\textrm{NL}}}$; and then the intensity $\lambda_{n}^{\textrm{NL}}$
will be easily acquired by taking a derivation of $\Lambda_{n}^{\textrm{NL}}$.
By using displacement theorem \cite{Blaszczyszyn13Using,Baccelli09Stochastic},
the point process $\overline{\Phi_{n}^{\textrm{NL}}}$ is Poisson
with intensity measure
\begin{align}
 & \quad\,\Lambda_{n}^{\textrm{NL}}\left(\left[0,t\right]\right)=\mathbb{E}_{\overline{\Phi_{n}^{\textrm{NL}}}}\left[\overline{\Phi_{n}^{\textrm{NL}}}\left[b\left(0,t\right)\right]\right]\nonumber \\
 & =\int_{\mathbb{R}^{2}}\Pr\left[\overline{R_{i,n}^{\textrm{NL}}}<t\right]p^{\textrm{NL}}\left(R_{i}\right)\lambda\textrm{d}\boldsymbol{X}_{i}\nonumber \\
 & =\mathbb{E}_{\mathcal{H}_{n}^{\textrm{NL}}}\left\{ \int_{\mathbb{R}^{2}}\Pr\left[R_{i}<t\left(B_{n}^{\textrm{NL}}\mathcal{H}_{n}^{\textrm{NL}}\right)^{1/\alpha_{n}^{\textrm{NL}}}\right]p_{n}^{\textrm{NL}}\left(R_{i}\right)\lambda\textrm{d}\boldsymbol{X}_{i}\right\} \nonumber \\
 & \overset{\left(a\right)}{=}\mathbb{E}_{\mathcal{H}_{n}^{\textrm{NL}}}\left[\int_{\theta=0}^{2\pi}\int_{R_{i}=0}^{t\left(B_{n}^{\textrm{NL}}\mathcal{H}_{n}^{\textrm{NL}}\right)^{1/\alpha_{n}^{\textrm{NL}}}}p_{n}^{\textrm{NL}}\left(R_{i}\right)\lambda R_{i}\textrm{d}R_{i}\textrm{d}\theta\right]\nonumber \\
 & =\mathbb{E}_{\mathcal{H}_{n}^{\textrm{NL}}}\left[2\pi\lambda\int_{R_{i}=0}^{t\left(B_{n}^{\textrm{NL}}\mathcal{H}_{n}^{\textrm{NL}}\right)^{1/\alpha_{n}^{\textrm{NL}}}}p_{n}^{\textrm{NL}}\left(R_{i}\right)R_{i}\textrm{d}R_{i}\right],\label{eq:Lambda_N proof-1}
\end{align}
where $b\left(0,t\right)$ is a ball centered at the origin $o$ with
radius $t$ and $\left(a\right)$ results by converting from Cartesian
to polar coordinates. Considering the distance range of $d_{n-1}<R_{i}\leqslant d_{n}$
(define $d_{0}$ and $d_{N}$ as 0 and $\infty$, respectively), the
equation above should be revised as follows
\begin{align}
\Lambda_{n}^{\textrm{NL}}\left(\left[0,t\right]\right) & =\mathbb{E}_{\mathcal{H}_{n}^{\textrm{NL}}}\left[2\pi\lambda\int_{R_{i}=d_{n-1}}^{R_{i,\max}^{\textrm{NL}}}p_{n}^{\textrm{NL}}\left(R_{i}\right)R_{i}\textrm{d}R_{i}\right],\label{eq:Lambda_N proof}
\end{align}
where we define $R_{i,\max}^{\textrm{NL}}=\min\left\{ d_{n},t\left(B_{n}^{\textrm{NL}}\mathcal{H}_{n}^{\textrm{NL}}\right)^{1/\alpha_{n}^{\textrm{NL}}}\right\} $.
Then the intensity of $\overline{\Phi_{n}^{\textrm{NL}}}$ denoted
by $\lambda_{n}^{\textrm{NL}}\left(\cdot\right)$ can be given by
\begin{equation}
\lambda_{n}^{\textrm{NL}}\left(t\right)=\frac{\textrm{d}}{\textrm{d}t}\Lambda_{n}^{\textrm{NL}}\left(\left[0,t\right]\right).
\end{equation}
 Note that to ensure the intensity measure is finite for any bounded
set (a set is bounded if it can be contained in a ball with a finite
radius), $\mathcal{H}_{n}^{\textrm{NL}}$ has to satisfy a certain
condition. As $p_{n}^{\textrm{NL}}\left(R_{i}\right)\leqslant1$,
from Eq. (\ref{eq:Lambda_N proof}), we get an inequality as follows
\begin{align}
 & \quad\,\Lambda_{n}^{\textrm{NL}}\left(\left[0,t\right]\right)=\mathbb{E}_{\mathcal{H}_{n}^{\textrm{NL}}}\left[2\pi\lambda\int_{d_{n-1}}^{R_{i,\max}^{\textrm{NL}}}p_{n}^{\textrm{NL}}\left(R_{i}\right)R_{i}\textrm{d}R_{i}\right]\nonumber \\
 & \leqslant\mathbb{E}_{\mathcal{H}_{n}^{\textrm{NL}}}\left[2\pi\lambda\int_{0}^{R_{i,\max}^{\textrm{NL}}}R_{i}\textrm{d}R_{i}\right]\nonumber \\
 & =\pi\lambda\min\left\{ d_{n}^{2},t^{2}\left(B_{n}^{\textrm{NL}}\right)^{2/\alpha_{n}^{\textrm{NL}}}\mathbb{E}_{\mathcal{H}_{n}^{NL}}\left[\left(\mathcal{H}_{n}^{\textrm{NL}}\right)^{2/\alpha_{n}^{\textrm{NL}}}\right]\right\} .
\end{align}
If the expectation $\mathbb{E}_{\mathcal{H}_{n}^{\textrm{NL}}}\left[\left(\mathcal{H}_{n}^{\textrm{NL}}\right)^{2/\alpha_{n}^{\textrm{NL}}}\right]<\infty$,
then $\Lambda_{n}^{\textrm{NL}}\left(\left[0,t\right]\right)<\infty$.
Using similar approach, the intensity measure and intensity of the
PPP $\overline{\Phi_{n}^{\textrm{L}}}$ are obtained by Eq. (\ref{eq:Measure_L})
and Eq. (\ref{eq:Lambda_L}), respectively.

As for the cell association scheme, it is obvious that the original
scheme $\left(\boldsymbol{X}_{i},\textrm{U},\mathcal{N}\right)^{*}=\arg\underset{(\boldsymbol{X}_{i},\textrm{U},\mathcal{N})\in\mathbb{S}}{\max}B_{n}^{\textrm{U}}\mathcal{H}_{n}^{\textrm{U}}\left(R_{i}\right)^{-\alpha_{n}^{\textrm{U}}}$
is equivalent to the scheme $\left(\boldsymbol{X}_{i},\textrm{U},\mathcal{N}\right)^{*}=\arg\underset{(\boldsymbol{X}_{i},\textrm{U},\mathcal{N})\in\mathbb{S}}{\max}\left(\overline{R_{i,n}}\right)^{-\alpha_{n}^{\textrm{U}}}$
which actually corresponds to the nearest BS association scheme. Thus
the proof is completed.

\section*{Appendix B: Proof of Lemma \ref{lem:CDF_strongest}}

Denote the strongest NLoS received signal power and the strongest
LoS received signal power by $\mathcal{P}^{\textrm{NL}}$ and $\mathcal{P}^{\textrm{L}}$,
respectively. Note that we drop subscript $n$ under this special
case for simplicity. That is, $\mathcal{P}^{\textrm{NL}}=\max\left(P_{i}^{\textrm{NL}}\right)$
and $\mathcal{P}^{\textrm{L}}=\max\left(P_{i}^{\textrm{L}}\right)$.
Then the probability $\Pr\left[\mathcal{P}\leqslant\gamma\right]$
can be derived as
\begin{align}
 & \quad\,\Pr\left[\mathcal{P}\leqslant\gamma\right]\nonumber \\
 & =\Pr\left[\max\left(\overline{R_{i}^{\textrm{NL}}}^{-\alpha^{\textrm{NL}}}\right)\leqslant\gamma\cap\max\left(\overline{R_{i}^{\textrm{L}}}^{-\alpha^{\textrm{L}}}\right)\leqslant\gamma\right]\nonumber \\
 & =\Pr\left[\min\left(\overline{R_{i}^{\textrm{NL}}}\right)\geqslant\gamma^{-1/\alpha^{\textrm{NL}}}\cap\min\left(\overline{R_{i}^{\textrm{L}}}\right)\geqslant\gamma^{-1/\alpha^{\textrm{L}}}\right]\nonumber \\
 & =\Pr\left[\textrm{no nodes within }\gamma^{-1/\alpha^{\textrm{NL}}}\cap\textrm{no nodes within }\gamma^{-1/\alpha^{\textrm{L}}}\right]\nonumber \\
 & =\Pr\left[\overline{\Phi^{\textrm{NL}}}\left(b\left(0,\gamma^{-1/\alpha^{\textrm{NL}}}\right)\right)=0\cap\overline{\Phi^{\textrm{L}}}\left(b\left(0,\gamma^{-1/\alpha^{\textrm{L}}}\right)\right)=0\right]\nonumber \\
 & \overset{\left(a\right)}{=}\Pr\left[\overline{\Phi^{\textrm{NL}}}\left(b\left(0,\gamma^{-1/\alpha^{\textrm{NL}}}\right)\right)=0\right]\nonumber \\
 & \quad\,\times\Pr\left[\overline{\Phi^{\textrm{L}}}\left(b\left(0,\gamma^{-1/\alpha^{\textrm{L}}}\right)\right)=0\right]\nonumber \\
 & \overset{\left(b\right)}{=}\exp\left[-\Lambda^{\textrm{NL}}\left(\left[0,\gamma^{-1/\alpha^{\textrm{NL}}}\right]\right)\right]\cdot\exp\left[-\Lambda^{\textrm{L}}\left(\left[0,\gamma^{-1/\alpha^{\textrm{L}}}\right]\right)\right],\label{eq:CDF_strongest power}
\end{align}
where the notation $\overline{\Phi^{\textrm{U}}}\left(\Xi\right)$
refers to the number of points $x\in\overline{\Phi^{\textrm{U}}}$
contained in the set $\Xi$, while equality $\left(a\right)$ follows
from the independence of PPP $\overline{\Phi^{\textrm{NL}}}$ and
PPP $\overline{\Phi^{\textrm{L}}}$ , and $\left(b\right)$ comes
from the fact that the void probability $\Pr\left[\overline{\Phi^{\textrm{U}}}\left(b\left(0,r\right)\right)=0\right]=\exp\left[-\Lambda^{\textrm{U}}\left(\left[0,r\right]\right)\right]$
for a non-homogeneous PPP. Then the rest of the proof is straightforward.

\section*{Appendix C: Proof of Theorem \ref{thm:Pcoverage }}

By invoking the law of total probability and considering the independence
between $\overline{\Phi_{n}^{\textrm{NL}}}$ and $\overline{\Phi_{n}^{\textrm{L}}}$,
the coverage probability can be divided into two parts in each segment,
i.e., $p_{c,n}^{\textrm{NL}}\left(\lambda,T\right)$ and $p_{c,n}^{\textrm{L}}\left(\lambda,T\right)$,
which denotes the conditional coverage probability given that the
typical MU is associated with a BS in $\Phi_{n}^{\textrm{NL}}$ and
$\Phi_{n}^{\textrm{L}}$, respectively. Moreover, denote by $\mathcal{P}_{n}^{\textrm{NL}}$
and $\mathcal{P}_{n}^{\textrm{L}}$ the strongest received signal
power from BS in $\Phi_{n}^{\textrm{NL}}$ and $\Phi_{n}^{\textrm{L}}$,
i.e., $\mathcal{P}_{n}^{\textrm{NL}}=\max\left(P_{i,n}^{\textrm{NL}}\right)$
and $\mathcal{P}_{n}^{\textrm{L}}=\max\left(P_{i,n}^{\textrm{L}}\right)$,
respectively. Then by applying the law of total probability, $p_{c,n}^{\textrm{L}}\left(\lambda,T\right)$
can be computed by
\begin{align}
p_{c,n}^{\textrm{L}}\left(\lambda,T\right) & =\Pr\left[\left(\textrm{SINR}_{n}^{\textrm{L}}>T\right)\cap\left(\mathcal{P}_{n}^{\textrm{L}}>\mathcal{P}_{n}^{\textrm{NL}}\right)\cap\mathcal{Y}_{n}^{\textrm{L}}\right]\nonumber \\
 & =\mathbb{E}_{\mathcal{Y}_{n}^{\textrm{L}}}\biggl\{\underset{\textrm{II}}{\underbrace{\Pr\left[\left.\textrm{SINR}_{n}^{\textrm{L}}>T\right|\left(\mathcal{P}_{n}^{\textrm{L}}>\mathcal{P}_{n}^{\textrm{NL}}\right)\cap\mathcal{Y}_{n}^{\textrm{L}}\right]}}\nonumber \\
 & \quad\,\times\underset{\textrm{I}}{\underbrace{\Pr\left[\left.\mathcal{P}_{n}^{\textrm{L}}>\mathcal{P}_{n}^{\textrm{NL}}\right|\mathcal{Y}_{n}^{\textrm{L}}\right]}}\biggr\},\label{eq:proof_pcL}
\end{align}
where $\mathcal{Y}_{n}^{\textrm{L}}$ is the equivalent distance between
the typical MU and the BS providing the strongest received signal
power to the typical MU in $\Phi_{n}^{\textrm{L}}$, i.e., $\mathcal{Y}_{n}^{\textrm{L}}=\underset{\overline{R_{i,n}^{\textrm{L}}}\in\overline{\Phi_{n}^{\textrm{L}}}}{\arg\max}\left(\overline{R_{i,n}^{\textrm{L}}}\right)^{-\alpha_{n}^{\textrm{L}}}$,
and also note that $\mathcal{P}_{n}^{\textrm{L}}=\left(\mathcal{Y}_{n}^{\textrm{L}}\right)^{-\alpha_{n}^{\textrm{L}}}$.
Besides, Part I guarantees that the typical MU is connected to a LoS
BS and Part II denotes the coverage probability conditioned on the
proposed cell association scheme in Eq. (\ref{eq:inspower}). Next,
Part I and Part II will be respectively derived as follows. For Part
I,
\begin{align}
 & \quad\,\Pr\left[\left.\mathcal{P}_{n}^{\textrm{L}}>\mathcal{P}_{n}^{\textrm{NL}}\right|\mathcal{Y}_{n}^{\textrm{L}}\right]\nonumber \\
 & =\Pr\left[\left.\left(\mathcal{Y}_{n}^{\textrm{L}}\right)^{-\alpha_{n}^{\textrm{L}}}>\left(\mathcal{Y}_{n}^{\textrm{NL}}\right)^{-\alpha_{n}^{\textrm{NL}}}\right|\mathcal{Y}_{n}^{\textrm{L}}\right]\nonumber \\
 & \overset{\left(a\right)}{=}\exp\left[-\Lambda_{n}^{\textrm{NL}}\left(\left[0,\left(\mathcal{Y}_{n}^{\textrm{L}}\right)^{\alpha_{n}^{\textrm{L}}/\alpha_{n}^{\textrm{NL}}}\right]\right)\right],\label{eq:proof_PL_g_PN}
\end{align}
where $\mathcal{Y}_{n}^{\textrm{NL}}$, similar to the definition
of $\mathcal{Y}_{n}^{\textrm{L}}$, is the equivalent distance between
the typical MU and the BS providing the strongest received signal
power to the typical MU in $\Phi_{n}^{\textrm{NL}}$, i.e., $\mathcal{Y}_{n}^{\textrm{NL}}=\underset{\overline{R_{i,n}^{\textrm{NL}}}\in\overline{\Phi_{n}^{\textrm{NL}}}}{\arg\max}\left(\overline{R_{i,n}^{\textrm{NL}}}\right)^{-\alpha_{n}^{\textrm{NL}}}$,
and also note that $\mathcal{P}_{n}^{\textrm{NL}}=\left(\mathcal{Y}_{n}^{\textrm{NL}}\right)^{-\alpha_{n}^{\textrm{NL}}}$,
and $\left(a\right)$ follows from the void probability of a PPP.

For Part II, we know that $\textrm{SINR}=\frac{\mathcal{P}}{I+\eta}=\frac{\mathcal{P}}{I^{\textrm{NL}}+I^{\textrm{L}}+\eta},$
where $I^{\textrm{NL}}$ and $I^{\textrm{L}}$ denote the aggregate
interference from NLoS BSs and LoS BSs, respectively. The conditional
coverage probability is derived as follows
\begin{align}
 & \quad\,\Pr\left[\left.\textrm{SINR}_{n}^{\textrm{L}}>T\right|\left(\mathcal{P}_{n}^{\textrm{L}}>\mathcal{P}_{n}^{\textrm{NL}}\right)\cap\mathcal{Y}_{n}^{\textrm{L}}\right]\nonumber \\
 & =\Pr\left[\left.\frac{1}{\textrm{SINR}_{n}^{\textrm{L}}}<\frac{1}{T}\right|\left(\mathcal{P}_{n}^{\textrm{L}}>\mathcal{P}_{n}^{\textrm{NL}}\right)\cap\mathcal{Y}_{n}^{\textrm{L}}\right]\nonumber \\
 & \overset{\left(a\right)}{=}\underset{\textrm{PDF}}{\int_{x=0}^{1/T}\underbrace{\int_{\omega=-\infty}^{\infty}\frac{e^{-j\omega x}}{2\pi}\mathcal{F}_{\frac{1}{\textrm{SINR}_{n}^{\textrm{L}}}}\left(\omega\right)\textrm{d}\omega}\textrm{d}x}\nonumber \\
 & =\int_{\omega=-\infty}^{\infty}\left[\frac{1-e^{-j\omega/T}}{2\pi j\omega}\right]\mathcal{F}_{\frac{1}{\textrm{SINR}_{n}^{\textrm{L}}}}\left(\omega\right)\textrm{d}\omega,\label{eq:proof_SINR}
\end{align}
where $\textrm{SINR}_{n}^{\textrm{L}}$ denotes the SINR when the
typical MU is associated with a LoS BS, the inner integral in $\left(a\right)$,
i.e., $\int_{\omega=-\infty}^{\infty}\frac{e^{-j\omega x}}{2\pi}\mathcal{F}_{\frac{1}{\textrm{SINR}_{n}^{\textrm{L}}}}\left(\omega\right)\textrm{d}\omega$
is the conditional PDF of $\frac{1}{\textrm{SINR}_{n}^{\textrm{L}}}$
due to the definition of the inverse characteristic function, i.e.,
$f_{X}\left(x\right)=F_{X}^{'}\left(x\right)=\frac{1}{2\pi}\int_{\mathrm{R}}e^{-j\omega x}\varphi_{X}\left(\omega\right)\textrm{d}\omega$,
and $\mathcal{F}_{\frac{1}{\textrm{SINR}_{n}^{\textrm{L}}}}\left(\omega\right)$
denotes the conditional characteristic function of $\frac{1}{\textrm{SINR}_{n}^{\textrm{L}}}$
which is given by
\begin{align}
 & \mathcal{F}_{\frac{1}{\textrm{SINR}_{n}^{\textrm{L}}}}\left(\omega\right)=\mathbb{E}_{\Phi_{n}}\left[\left.\exp\left(j\omega\frac{1}{\textrm{SINR}_{n}^{\textrm{L}}}\right)\right|\left(\mathcal{P}_{n}^{\textrm{L}}>\mathcal{P}_{n}^{\textrm{NL}}\right)\cap\mathcal{Y}_{n}^{\textrm{L}}\right]\nonumber \\
 & =\mathbb{E}_{\Phi_{n}}\left[\left.\exp\left(j\omega\frac{I^{\textrm{NL}}+I^{\textrm{L}}+\eta}{\mathcal{P}_{n}^{\textrm{L}}}\right)\right|\left(\mathcal{P}_{n}^{\textrm{L}}>\mathcal{P}_{n}^{\textrm{NL}}\right)\cap\mathcal{Y}_{n}^{\textrm{L}}\right]\nonumber \\
 & =\mathbb{E}_{\Phi_{n}}\left\{ \left.\exp\left[j\omega\left(I^{\textrm{NL}}+I^{\textrm{L}}+\eta\right)\left(\mathcal{Y}_{n}^{\textrm{L}}\right)^{\alpha_{n}^{\textrm{L}}}\right]\right|\left(\mathcal{P}_{n}^{\textrm{L}}>\mathcal{P}_{n}^{\textrm{NL}}\right)\cap\mathcal{Y}_{n}^{\textrm{L}}\right\} \nonumber \\
 & \overset{\left(a\right)}{=}\underset{\textrm{III}}{\underbrace{\mathbb{E}_{\Phi_{n}^{\textrm{NL}}}\left\{ \left.\exp\left[j\omega I^{\textrm{NL}}\cdot\left(\mathcal{Y}_{n}^{\textrm{L}}\right)^{\alpha_{n}^{\textrm{L}}}\right]\right|\left(\mathcal{P}_{n}^{\textrm{L}}>\mathcal{P}_{n}^{\textrm{NL}}\right)\cap\mathcal{Y}_{n}^{\textrm{L}}\right\} }}\nonumber \\
 & \quad\,\times\underset{\textrm{IV}}{\underbrace{\mathbb{E}_{\Phi_{n}^{\textrm{L}}}\left\{ \left.\exp\left[j\omega I^{\textrm{L}}\cdot\left(\mathcal{Y}_{n}^{\textrm{L}}\right)^{\alpha_{n}^{\textrm{L}}}\right]\right|\left(\mathcal{P}_{n}^{\textrm{L}}>\mathcal{P}_{n}^{\textrm{NL}}\right)\cap\mathcal{Y}_{n}^{\textrm{L}}\right\} }}\nonumber \\
 & \quad\,\times e^{j\omega\eta\left(\mathcal{Y}_{n}^{\textrm{L}}\right)^{\alpha_{n}^{\textrm{L}}}},\label{eq:Characteristic_1/SINR}
\end{align}
where $\left(a\right)$ comes from the facts that $\Phi_{n}=\Phi_{n}^{\textrm{NL}}\cup\Phi_{n}^{\textrm{L}}$
and the mutual independence of $\Phi_{n}^{\textrm{NL}}$ and $\Phi_{n}^{\textrm{L}}$.
Now by applying stochastic geometry, we will derive the term III in
Eq. (\ref{eq:Characteristic_1/SINR}) as follows
\begin{align}
 & \quad\,\mathbb{E}_{\Phi_{n}^{\textrm{NL}}}\left\{ \left.\exp\left[j\omega I^{\textrm{NL}}\cdot\left(\mathcal{Y}_{n}^{\textrm{L}}\right)^{\alpha_{n}^{\textrm{L}}}\right]\right|\left(\mathcal{P}_{n}^{\textrm{L}}>\mathcal{P}_{n}^{\textrm{NL}}\right)\cap\mathcal{Y}_{n}^{\textrm{L}}\right\} \nonumber \\
 & \overset{\left(a\right)}{=}\mathbb{E}_{\overline{\Phi_{n}^{\textrm{NL}}}}\Biggl\{\left.\exp\left[j\omega\cdot\left(\mathcal{Y}_{n}^{\textrm{L}}\right)^{\alpha_{n}^{\textrm{L}}}\underset{i:\overline{R_{i,n}^{\textrm{NL}}}\in\overline{\Phi_{n}^{\textrm{NL}}}'}{\sum}\left(\overline{R_{i,n}^{\textrm{NL}}}\right)^{-\alpha_{n}^{\textrm{NL}}}\right]\right|\bigl(\mathcal{P}_{n}^{\textrm{L}}\nonumber \\
 & \quad\,>\mathcal{P}_{n}^{\textrm{NL}}\bigr)\cap\mathcal{Y}_{n}^{\textrm{L}}\Biggr\}\nonumber \\
 & \overset{\left(b\right)}{=}\mathbb{E}_{\overline{\Phi_{n}^{\textrm{NL}}}}\Biggl\{\left.\underset{i:\overline{R_{i,n}^{\textrm{NL}}}\in\overline{\Phi_{n}^{\textrm{NL}}}'}{\prod}\exp\left[j\omega\cdot\left(\mathcal{Y}_{n}^{\textrm{L}}\right)^{\alpha_{n}^{\textrm{L}}}\left(\overline{R_{i,n}^{\textrm{NL}}}\right)^{-\alpha_{n}^{\textrm{NL}}}\right]\right|\bigl(\mathcal{P}_{n}^{\textrm{L}}\nonumber \\
 & \quad\,>\mathcal{P}_{n}^{\textrm{NL}}\bigr)\cap\mathcal{Y}_{n}^{\textrm{L}}\Biggr\}\nonumber \\
 & \overset{\left(c\right)}{=}\exp\left\{ \int_{t=\left(\mathcal{Y}_{n}^{\textrm{L}}\right)^{\alpha_{n}^{\textrm{L}}/\alpha_{n}^{\textrm{NL}}}}^{\infty}\left[e^{j\omega\left(\mathcal{Y}_{n}^{\textrm{L}}\right)^{\alpha_{n}^{\textrm{L}}}t^{-\alpha_{n}^{\textrm{NL}}}}-1\right]\lambda_{n}^{\textrm{NL}}\left(t\right)\textrm{d}t\right\} ,\label{eq:proof_E_N}
\end{align}
where in $\left(a\right)$, $\overline{\Phi_{n}^{\textrm{NL}}}'=\overline{\Phi_{n}^{\textrm{NL}}}\setminus b\left(0,\left(\mathcal{Y}_{n}^{\textrm{L}}\right)^{\alpha_{n}^{\textrm{L}}/\alpha_{n}^{\textrm{NL}}}\right)$
and $\overline{R_{i,n}^{\textrm{NL}}}\in\overline{\Phi_{n}^{\textrm{NL}}}'$
can guarantee the condition that $\mathcal{P}_{n}^{\textrm{L}}>\mathcal{P}_{n}^{\textrm{NL}}$,
$\left(b\right)$ follows from rewriting the exponential of summation
as a product of several exponential functions, and $\left(c\right)$
is obtained by applying the probability generating functional (PGFL)
\cite[Eq. (3)]{Andrews11A} of the PPP. Similarly, the term IV in
Eq. (\ref{eq:Characteristic_1/SINR}) is given by
\begin{align}
 & \quad\,\mathbb{E}_{\Phi_{n}^{\textrm{L}}}\left\{ \left.\exp\left[j\omega I^{\textrm{L}}\cdot\left(\mathcal{Y}_{n}^{\textrm{L}}\right)^{\alpha_{n}^{\textrm{L}}}\right]\right|\left(\mathcal{P}_{n}^{\textrm{L}}>\mathcal{P}_{n}^{\textrm{NL}}\right)\cap\mathcal{Y}_{n}^{\textrm{L}}\right\} \nonumber \\
 & \overset{\left(a\right)}{=}\mathbb{E}_{\overline{\Phi_{n}^{\textrm{L}}}}\Biggl\{\left.\exp\left[j\omega\cdot\left(\mathcal{Y}_{n}^{\textrm{L}}\right)^{\alpha_{n}^{\textrm{L}}}\underset{i:\overline{R_{i,n}^{\textrm{L}}}\in\overline{\Phi_{n}^{\textrm{L}}}'}{\sum}\left(\overline{R_{i,n}^{\textrm{L}}}\right)^{-\alpha_{n}^{\textrm{L}}}\right]\right|\bigl(\mathcal{P}_{n}^{\textrm{L}}\nonumber \\
 & \quad\,>\mathcal{P}_{n}^{\textrm{NL}}\bigr)\cap\mathcal{Y}_{n}^{\textrm{L}}\Biggr\}\nonumber \\
 & =\mathbb{E}_{\overline{\Phi_{n}^{\textrm{NL}}}}\Biggl\{\left.\underset{i:\overline{R_{i,n}^{\textrm{L}}}\in\overline{\Phi_{n}^{\textrm{L}}}'}{\prod}\exp\left[j\omega\cdot\left(\mathcal{Y}_{n}^{\textrm{L}}/\overline{R_{i,n}^{\textrm{L}}}\right)^{\alpha_{n}^{\textrm{L}}}\right]\right|\bigl(\mathcal{P}_{n}^{\textrm{L}}\nonumber \\
 & \quad\,>\mathcal{P}_{n}^{\textrm{NL}}\bigr)\cap\mathcal{Y}_{n}^{\textrm{L}}\Biggr\}\nonumber \\
 & =\exp\left\{ \int_{t=\mathcal{Y}_{n}^{\textrm{L}}}^{\infty}\left[e^{j\omega\left(\mathcal{Y}_{n}^{\textrm{L}}/t\right)^{\alpha_{n}^{\textrm{L}}}}-1\right]\lambda_{n}^{\textrm{L}}\left(t\right)\textrm{d}t\right\} ,\label{eq:proof_E_L}
\end{align}
where in $\left(a\right)$, $\overline{\Phi_{n}^{\textrm{L}}}'=\overline{\Phi_{n}^{\textrm{L}}}\setminus b\left(0,\mathcal{Y}_{n}^{\textrm{L}}\right)$
and $\overline{R_{i,n}^{\textrm{L}}}\in\overline{\Phi_{n}^{\textrm{L}}}'$
can guarantee that the typical MU is associated with a LoS BS providing
the strongest received signal power. Then the product of Part I and
Part II in Eq. (\ref{eq:proof_pcL}) can be obtained by substituting
them with Eq. (\ref{eq:proof_PL_g_PN}) \textendash{} (\ref{eq:proof_E_L}).

Finally, note that the value of $p_{c,n}^{\textrm{L}}\left(\lambda,T\right)$
in Eq. (\ref{eq:proof_pcL}) should be calculated by taking the expectation
with respect to $\mathcal{Y}_{n}^{\textrm{L}}$ in terms of its PDF,
which is given as follows
\begin{align}
f_{\mathcal{Y}_{n}^{\textrm{L}}}\left(y\right)=\frac{\textrm{d}}{\textrm{d}y}\left[1-\Pr\left(\mathcal{Y}_{n}^{\textrm{L}}>y\right)\right] & =\lambda_{n}^{\textrm{L}}\left(y\right)\exp\left[-\Lambda_{n}^{\textrm{L}}\left(\left[0,y\right]\right)\right].\label{eq:pdf_y_N}
\end{align}

Given that the typical MU is connected to a NLoS BS, the conditional
coverage probability $p_{c,n}^{\textrm{NL}}\left(\lambda,T\right)$
can be derived in a similar way as the above. In this way, the coverage
probability is obtained by $p_{c}\left(\lambda,T\right)=\stackrel[n=1]{N}{\sum}p_{c,n}^{\textrm{L}}\left(\lambda,T\right)+\stackrel[n=1]{N}{\sum}p_{c,n}^{\textrm{NL}}\left(\lambda,T\right)$.
Thus the proof is completed.

\bibliographystyle{IEEEtran}
\bibliography{BY}

\end{document}